\documentclass[bibyear]{aa} 

\usepackage{natbib}
\bibpunct{(}{)}{;}{a}{}{,} 
\makeatletter
\renewcommand*\aa@pageof{, page \thepage{} of \pageref*{LastPage}}
\makeatother
\usepackage{graphicx}
\usepackage{txfonts}
\usepackage[colorlinks]{hyperref}
\hypersetup{
    citecolor = {cyan},
    linkcolor = {magenta}}
\begin{document}

   \title{Diagnosing the interstellar medium of galaxies with far-infrared emission lines}

   \subtitle{I. The {[\ion{C}{II}]} 158 $\mu$m line at z$\sim$0}

   \author{A. F. Ramos Padilla \thanks{\email{ramos@astro.rug.nl}} \inst{1,2} 
          \and
        L. Wang\inst{1,2} 
        \and
        S. Ploeckinger
        \inst{3}
        \and
        F. F. S. van der Tak
        \inst{1,2}
        \and 
        S. C. Trager\inst{1}
          }

   \institute{Kapteyn Astronomical Institute, University of Groningen, Landleven 12, 9747 AD, Groningen, the Netherlands 
         \and
             SRON Netherlands Institute for Space Research, Landleven 12, 9747 AD, Groningen, the Netherlands
        \and
            Lorentz Institute for Theoretical Physics, Leiden University, 2333 CA Leiden, the Netherlands  \\
             }

   \date{Received September 15, 1996; accepted March 16, 1997}

 
  \abstract
   {Atomic fine structure lines have been detected in the local Universe and at high redshifts over the past decades. The {[\ion{C}{II}]} emission line at 158 $\mu$m is an important observable as it provides constraints on the interstellar medium (ISM) cooling processes.}
   {We develop a physically motivated framework to simulate the production of far-infrared line emission from galaxies in a cosmological context. This first paper sets out our methodology and describes its first application: simulating the {[\ion{C}{II}]} 158 $\mu$m line emission in the local Universe.}
   {We combine the output from \textsc{EAGLE} cosmological hydrodynamical simulations with a multi-phase model of the ISM. Gas particles are divided into three phases: dense molecular gas, neutral atomic gas, and diffuse ionised gas (DIG). We estimate the {[\ion{C}{II}]} line emission from the three phases using a set of \textsc{Cloudy} cooling tables.}
   {Our results agree with previous findings regarding the contribution of these three ISM phases to the {[\ion{C}{II}]} emission. Our model shows good agreement with the observed $L_{\mathrm[\ion{C}{II}]}$--star formation rate (SFR) relation in the local Universe within 0.4 dex scatter.}
  {The fractional contribution to the {[\ion{C}{II}]} line from different ISM phases depends on the total SFR and metallicity. The neutral gas phase dominates the {[\ion{C}{II}]} emission in galaxies with $\mathrm{SFR}\sim0.01$--$1\,\mathrm{M_{\sun}}\,\mathrm{yr^{-1}}$, but the ionised phase dominates at lower SFRs. Galaxies above solar metallicity exhibit lower $L_{\mathrm[\ion{C}{II}]}$/SFR ratios for the neutral phase. In comparison, the $L_{\mathrm[\ion{C}{II}]}$/SFR ratio in the DIG is stable when metallicity varies. We suggest that the reduced size of the neutral clouds, caused by increased SFRs, is the likely cause for the $L_{\mathrm[\ion{C}{II}]}$ deficit at high infrared luminosities, although \textsc{EAGLE}  simulations do not reach these luminosities at $z=0$.}


   \maketitle
%

\section{Introduction}\label{sec:intro}

Over the past decades, significant efforts to develop theoretical models have helped improve our knowledge of how galaxies form and evolve. Implementations of hydrodynamical simulations \citep[e.g.][]{2014MNRAS.444.1518V,2014MNRAS.445..581H,2015MNRAS.446..521S,2018MNRAS.473.4077P} and semi-analytic models \citep[e.g.][]{1999MNRAS.310.1087S,2000MNRAS.319..168C,2006MNRAS.365...11C} have provided fundamental insights. Most models can now roughly reproduce the observed specific star formation rate (SFR) and mimic the scenario where more massive galaxies formed their stars earlier than lower mass galaxies, an effect known as `downsizing' \citep{1996AJ....112..839C,2008ApJ...675..234P,2017A&A...605A...4H}. These implementations agree, within a factor of three, with physical properties of galaxies and, in recent years, have begun to converge regarding physical interpretations \citep{2015ARA&A..53...51S}.

One critical question, tackled by several works in recent years \citep[see e.g.][]{2006ApJ...647...60N,2013MNRAS.433.1567V,2015ApJ...813...36V,2017MNRAS.468.4831K,2019MNRAS.487.5902K,2017MNRAS.471.4128P,2017MNRAS.465.2540P,2015ApJ...814...76O,2017ApJ...846..105O,2020MNRAS.492.2818L,2020MNRAS.496.5160L,2018MNRAS.481L..84M,2020MNRAS.498.5541A}, is how to include the cooling processes of the interstellar medium (ISM) in galactic theoretical models beginning from early epochs. Unfortunately, there is still no single model capable of reproducing all the observational data across all of cosmic history. The balance of gas heating and cooling in the ISM environment (which affects physical properties such as temperature and density) is central to theoretical models.
Owing to observational constraints, our ISM knowledge mainly comes from our Galaxy and its nearby cosmic neighbourhood, where cold atomic clouds (cold neutral medium, CNM), diffuse warm neutral and ionised emission (WNM and WIM), and \ion{H}{II} regions are distinguishable  \citep[e.g.][]{ 1999RvMP...71..173H,1995ApJ...443..152W,2003ApJ...587..278W,1999ApJ...527..795K}.

Far-infrared (FIR) emission line observations of nearby galaxies that trace these phases of the ISM \citep{2017ApJ...846...32D,2018ApJ...861...95H} are an important tool for understanding ISM cooling processes and their relation with the SFR, especially in cool gas where permitted lines of hydrogen cannot be excited ($<10^{4}$ K). However, the current resolution of (most) instruments limits the capability to provide spatially resolved line observations of high-redshift galaxies as in nearby galaxies. There are a few exceptions: Gas an dust clouds can be observed at high redshifts using high spatial resolution with ALMA \citep[e.g.][]{2017ApJ...837..182O} or using gravitational lensing \citep{2019NatAs...3.1114D}. Extrapolating observations from the local Universe is not a good option as the ISM phases are likely to be different at earlier times. For example, gas-phase metallicities  change with redshift, which will have an impact on the ISM phases \citep{2019ApJ...881..160B}. The growing body of FIR line observations at high-redshift has so far mainly been interpreted based on empirical relations. For example, the $L_{\mathrm[\ion{C}{II}]}$--SFR relations obtained by \citet{2014A&A...568A..62D} are used to interpret high-redshift galaxies \citep{2016Sci...352.1559I, 2016ApJ...829L..11P,2016MNRAS.462L...6K}. However, over the last few years, new high redshift observations have begun to use emission models to interpret the observations of these lines \citep[e.g.][]{2015MNRAS.452...54M,2017A&A...605A..42C,2020MNRAS.493.4294B,2020arXiv200200962B}.

Hydrodynamical simulations represent one of the most promising methods to avoid this extrapolation from the local Universe to high redshifts. These simulations can predict the interplay between dark matter and baryons in the large-scale structure and the final properties of galaxies \citep{2018PhR...780....1D}. However, this method is computationally expensive if all the relevant physics are considered. Limitations in spatial resolutions and simulation techniques have to be taken into account in the sub-grid physics in different box sizes \citep[see][their Table~1]{2018PhR...780....1D}. Fortunately, zoom-in techniques are starting to bridge the gap between individual stellar and galactic scales, which help to model the general ISM \citep{2015ARA&A..53...51S}. Therefore, hydrodynamical simulations can help us to predict emission lines at different cosmic epochs and characterise the ISM at different cosmic times \citep{2017MNRAS.471.4128P}. 

The {[\ion{C}{II}]} 158 $\mu $m emission line is one of the brightest emission lines in the FIR. Its luminosity is equivalent to values around 1\% of the FIR luminosity of galaxies \citep[e.g.][]{1991ApJ...373..423S,2008ApJS..178..280B}. It is an easily observed line that traces various phases of the ISM where gas is exposed to energies above the carbon ionisation potential (11.3 eV compared to 13.6 eV for hydrogen). {[\ion{C}{II}]} can be considered as a robust cooling line (in the range of 20-8\,000 K) of the ISM, acting as a thermostat  \citep{2012ApJ...745...49G,2012ApJS..203...13G,2015ApJ...814...76O}. 

Nevertheless, {[\ion{C}{II}]} is difficult to interpret, as it arises from diverse environments: CNM, WNM and WIM. In addition, at higher FIR luminosities, the luminosity of {[\ion{C}{II}]} over the FIR luminosity increases at a lower rate, an effect known as the `{[\ion{C}{II}]} deficit' \citep[e.g.][who also observed deficits in other emission lines]{2017ApJ...846...32D}. These problems do not limit the capacity of {[\ion{C}{II}]} for tracing SFR in local luminous infrared galaxies \citep[e.g.][]{2001ApJ...561..766M,2010ApJ...724..957S,2013ApJ...774...68D, 2015ApJ...800....1H,2017ApJ...846...32D}, but a complete understanding of the origin of the deficit is necessary in order to use this line as a SFR indicator in high-redshift galaxies \citep{2011MNRAS.416.2712D,2015MNRAS.449.2883G,2016ApJ...826..112S}. Currently, dependencies on metallicities and radiation fields (radiative feedback) seem to be the most probable regulators of the {[\ion{C}{II}]} line luminosity in theoretical \citep[e.g.][]{2001ApJ...561..766M,2016MNRAS.463.2085M,2017MNRAS.467...50N,2017MNRAS.467.1300V,2019MNRAS.489....1F} and observational \citep{2010ApJ...724..957S,2017ApJ...846...32D,2018ApJ...861...95H} studies. 

The goal of this paper is to present a model for the ISM {[\ion{C}{II}]} line emission and show its applications in the local Universe ($z=0$). We implement this model of FIR line emission to comprehend the ISM physical conditions in galaxies with line properties. We use $z=0$ as a benchmark, as locally we have the best observational constraints that cover a diverse range of galaxies in different environments. Our first target is the {[\ion{C}{II}]} 158 $\mu$m  line at $z=0$, for which we assume contributions from the atomic, molecular, and ionised ISM phases to obtain a model of the {[\ion{C}{II}]} emission in galaxies. We model the emission of {[\ion{C}{II}]} by post-processing hydrodynamical simulations from the Evolution and Assembly of GaLaxies and their Environments (\textsc{EAGLE}) project \citep{2015MNRAS.446..521S,2015MNRAS.450.1937C} with a physically motivated model of the ISM. We use \textsc{Cloudy} \citep{1998PASP..110..761F,2013RMxAA..49..137F,2017RMxAA..53..385F} cooling tables \citep{2020MNRAS.497.4857P} to predict line emission to help us constrain different ISM phases in galactic environments. Throughout this paper, we assume the $\Lambda$CDM cosmology from Planck results \citep{2014A&A...571A..16P} ($\Omega= 0.307$, $\Omega_\Lambda$=0.693, $H_0 =67.7$ km s$^{-1}$ Mpc$^{-1}$ and $\sigma_8=0.8288$).

In this paper, we first give an overview of the methods we use to predict the emission lines (Sect.~\ref{sec:meth}). Then, we verify that our results agree with observations of local galaxies in terms of physical parameters and scaling relations (Sect.~\ref{sec:results}). After that, we discuss the difference between our findings and other papers. (Sect.~\ref{sec:disc}), and finally, we present our conclusions (Sect.~\ref{sec:conclu}).

\section{Methodology} \label{sec:meth}

In the next sections, we describe the sets of simulations we use (Sect.~\ref{subsec:EAGLEDes}), the model to characterise the multi-phase structure of the ISM (Sect.~\ref{subsec:ISMModel}), and the estimation of line luminosity using \textsc{Cloudy} cooling tables (\ref{subsec:CloudyGrid}). 

\subsection{The EAGLE simulations} \label{subsec:EAGLEDes}

\textsc{EAGLE} \citep{2015MNRAS.446..521S,2015MNRAS.450.1937C} consists of several cosmological hydrodynamical simulations, run in an N-body smoothed particle hydrodynamics (SPH) code. Briefly, \textsc{EAGLE} adopts a pressure-entropy parameterisation 
using the description of \citet{2013MNRAS.428.2840H}. The simulations include radiative cooling and photo-electric heating \citep{2009MNRAS.393...99W}, star formation \citep{2008MNRAS.383.1210S}, stellar evolution and
mass loss \citep{2009MNRAS.399..574W}, black hole growth \citep{2005MNRAS.361..776S,2015MNRAS.454.1038R}, and feedback from star formation and AGNs \citep{2012MNRAS.426..140D}. 

This work uses three simulations of different sizes and resolutions from \textsc{EAGLE}: \textsc{Ref-L0025N0376}, \textsc{Ref-L0100N1504}, and \textsc{Recal-L0025N0752}. Table~\ref{tab:EAGLESims} presents their main characteristics. We use these simulations to compare the implemented model when using different box-sizes and resolutions. Variations in terms of box-size are analysed when comparing the \textsc{Ref-L0100N1504} and \textsc{Ref-L0025N0376} simulations. Variations in terms of resolution are analysed when comparing the \textsc{Ref-L0025N0376} and \textsc{Recal-L0025N0752}. EAGLE simulations calibrate the physical parameters of the sub-grid routines to reproduce the observed galaxy stellar mass functions at $z \approx 0$ \citep[GSMF;][]{2015MNRAS.446..521S}. The sub-grid parameters of the high-resolution, small-box simulation \textsc{Recal-L0025N0752} were re-calibrated to account for the increased resolution \citep{2015MNRAS.446..521S,2015MNRAS.450.1937C}. \textsc{Ref-L0100N1504} and \textsc{Recal-L0025N0752} are similar in terms of `weak convergence', which means numerical results converge in different simulations after re-calibrating the sub-grid parameters \citep{2015MNRAS.450.4486F,2015MNRAS.446..521S}.

\begin{table*}
    \centering
    \caption{\textsc{EAGLE} simulations used in this work. The top simulation is the only high-resolution simulation used, while the other two are intermediate-resolution simulations with different box-sizes. The box-size (L) and maximum softening length are presented in comoving and proper distances (cMpc and pkpc). The last column shows the number of galaxies used in this work for a given simulation.}
    \label{tab:EAGLESims}
    \begin{tabular}{lcccccc}
    \hline
    \hline
    Name in & L & \# particles & Gas mass & Max. Softening & \# galaxies\\
    \citet{2015MNRAS.446..521S}&(cMpc)& & ($\mathrm{M_{\sun}}$) & (pkpc)&\\
    \hline
    \textsc{Recal-L0025N0752}&25& $752^{3}$ & $ 2.26\times 10^5$ & 0.35&415\\
    \textsc{Ref-L0025N0376} &25&$376^{3}$& $1.81 \times 10^6$ &0.7&202\\
    \textsc{Ref-L0100N1504} & 100 & $1\,504^{3}$&$1.81 \times 10^6$&0.7&5\,000\tablefootmark{a}\\
    \hline
    \end{tabular} 
    \tablefoot{
\tablefoottext{a}{We selected the top 5\,000 galaxies in terms of gas mass.}
}
\end{table*}

We selected our sample of galaxies from the \textsc{EAGLE} database \citep{2016A&C....15...72M} similarly to the study of \citet{2019MNRAS.485..972T}. We focus on `central' galaxies, sub-halos containing the particle with the lowest value of the gravitational potential, instead of  `satellite' galaxies, to avoid environmental influence on the morphology and kinematics of the galaxies. We use galaxies with at least 300 star particles within 30 pkpc (proper kpc) from the centre of the potential. For \textsc{Ref-L0100N1504}, the top 5\,000 most-gas-massive galaxies are selected to cover the range of gas mass that \textsc{Ref-L0025N0376} cannot cover. Simulated galaxies are retrieved from the SPH data \citep{2017arXiv170609899T} by using \textsc{FoF} (Friends-of-Friends) and \textsc{SUBFIND} algorithms \citep{2001MNRAS.328..726S,2009MNRAS.399..497D} in the dark matter halos. In the last column of Table~\ref{tab:EAGLESims}, we present the total number of galaxies used to calculate line luminosities with our model in each of the simulations.

\subsection{The multi-phase ISM model} \label{subsec:ISMModel}

In this section, we describe the model for the multi-phase ISM using gas and star SPH particles in terms of the simulation properties (Sect.~\ref{subsec:SPHModel}) and three different ISM environments: dense molecular clouds (Sect.~\ref{subsec:PDRMCModel}), neutral atomic gas (Sect.~\ref{subsec:PDRMCModel}), and diffuse ionised gas (DIG, Sect.~\ref{subsec:DIGModel}). 

\subsubsection{Gas and star particle properties}\label{subsec:SPHModel}

We selected SPH particles inside a volume of radius 30 pkpc to ensure that the derived parameters in our modelling are similar to those in the \textsc{EAGLE} database\footnote{\url{http://virgodb.dur.ac.uk/}} for each galaxy. In Table~\ref{tab:SPHprop}, we present the symbol and the properties from the particle data that we use for the modelling. In addition, we obtained the IDs for the particles and their position to study their spatial distribution. 

\begin{table*}
    \centering
    \caption{Physical properties used from the \textsc{EAGLE} data for stars and gas particles.}
    \label{tab:SPHprop}
    \begin{tabular}{lcl}
    \hline
    \hline
    Property & Symbol & Description [Units]\\
    \hline
    \multicolumn{3}{c}{Gas properties}\\
    \hline
    Density& $\rho$ & Co-moving density [g cm$^{-3}$]\\
    Smoothed element abundance hydrogen& $X_H$ & SPH weighted hydrogen abundance\\
    Smoothed element abundance carbon & $X_C$ & SPH weighted carbon abundance\\
    Entropy & $E$ & Particle entropy [Pressure $\times$ Density$^{\gamma}$]\\
    Mass & $m_{\mathrm{SPH}}$& Particle mass [g]\\
    Smoothed metallicity & $Z$ & SPH kernel weighted metallicity\\
    Smoothing length &$h$& Co-moving SPH smoothing kernel [cm]\\
    Star formation rate & SFR & Instantaneous star formation rate [M$_\sun$/yr]\\
    Temperature & $T_{\mathrm{SPH}}$ & Particle temperature (from the internal energy) [K]\\
    \hline
    \multicolumn{3}{c}{Star properties}\\
    \hline
    Star formation time& $t_{*}$ & Time when a star particle was born \\
    Stellar mass& $m_{*}$ & Current particle mass [g]\\
    Smoothed metallicity& $Z_{*}$ & Co-moving SPH smoothing metallicity\\
    \hline
    \end{tabular}    
\end{table*}

We retrieved physical parameters from the gas in each of the SPH particles before dividing these SPH into neutral and ionised gas phases. We calculated the total hydrogen number density as 
\begin{equation}
    n(\mathrm{H})=\frac{\rho X_H }{m_{H}},
\end{equation}
with $m_{H}$ the hydrogen mass, $\rho$ the density and $X_{H}$ the SPH weighted hydrogen abundance. \textsc{EAGLE} imposes a pressure floor expressed as a polytropic equation-of-state as $P \geq P_\mathrm{lim} \left(\rho/\rho_\mathrm{lim} \right)^{\gamma_{\mathrm{lim}}}$, where $\gamma_{\mathrm{lim}}=4/3$, $P_\mathrm{lim}$, and $\rho_\mathrm{lim}$ are constants  \citep[$\sim1.2 \times 10^{-13} \rm{Ba} \, \rm{and} \, \sim2.23 \times 10^{-25} \rm{g \, cm}^{-3}$,][]{2008MNRAS.383.1210S,2017arXiv170609899T}. A temperature threshold is then obtained from $P_\mathrm{lim} = \rho_\mathrm{lim}T_\mathrm{lim}/(\mu m_{H})$, where $\mu \approx 1.23$ is the mean molecular weight of the neutral gas. For gas particles limited by the pressure floor, the temperature $T_{\mathrm{lim}}$ is an effective temperature, including unresolved processes in addition to the thermal temperature. We therefore constrain $T_{\mathrm{SPH}}$ at $n(\mathrm{H})>0.1\, \mathrm{cm}^{-3}$ to 8\,000K, which is typical of the warm ISM (WNM and WIM). 

We follow \citet{2013MNRAS.430.2427R} to calculate the fraction of neutral hydrogen $\eta = n(H_{I})/n(\mathrm{H})$ given by the ionisation equilibrium $n(\ion{H}{I}) \Gamma_{\mathrm{Tot}} = \alpha_A n_e n(\ion{H}{II})$ as
\begin{equation}
    \displaystyle \Gamma_{\mathrm{Tot}} = \frac{\alpha_A \left( 1 - \eta \right)^2 n(\mathrm{H})}{\eta},
    \label{eq:IonEqui}
\end{equation}
where $\alpha_A$ is the Case A recombination rate, $\Gamma_{\mathrm{Tot}}$ is the total ionisation rate, and $n_e$, $n(\ion{H}{I})$, and $n(\ion{H}{II})$ are the number densities of electrons, neutral hydrogen and ionised hydrogen, respectively. To solve this equilibrium, we use $\alpha_A$ from \citet{1997MNRAS.292...27H},
\begin{equation}
    \displaystyle \alpha_A = 1.269 \times 10^{-13}  \frac{\lambda^{1.503}}{\left[ 1 + \left( \frac{\lambda}{0.522} \right)^{0.47}\right]^{1.923}} \mathrm{cm}^{3} \mathrm{s}^{-1},
    \label{eq:CaseA}
\end{equation}
where $\lambda=2T_{\mathrm{TR}}/T$, with $T_{\mathrm{TR}}=157\,807$K, is the ionisation threshold for \ion{H}{I}, and $T$ is the gas temperature. $\Gamma_{\mathrm{Tot}}$ can be defined as
\begin{equation}
    \Gamma_{\mathrm{Tot}}= \Gamma_{\mathrm{Phot}} + \Gamma_{\mathrm{Col}},
\end{equation} 
where $\Gamma_{\mathrm{Phot}}$ is the photo-ionisation rate, and $\Gamma_{\mathrm{Col}}$
is the collisional ionisation rate. $\Gamma_{\mathrm{Col}}$ is calculated following \citet{1998MNRAS.301..478T} as
\begin{equation}
    \Gamma_{\mathrm{Col}} = \Lambda_T (1 - \eta) n(\mathrm{H}),
\end{equation}
where
\begin{equation}
    \Lambda_T = 1.17 \times 10^{-10}
    \frac{T^{1/2} \exp{(-157\,809/T)}}{1 + \sqrt{T/10^5}} \mathrm{cm}^{3} \mathrm{s}^{-1}.
    \label{eq:ColRate}
\end{equation}

\noindent Rearranging Eq.~\ref{eq:IonEqui} with Eqs.~\ref{eq:CaseA}-\ref{eq:ColRate} as
\begin{eqnarray}
    \displaystyle A&=& \alpha_A + \Lambda_T \nonumber\\
    \displaystyle B&=& 2 \alpha_A + \frac{\Gamma_{\mathrm{Phot}}}{n(\mathrm{H})} + \Lambda_T \nonumber \\
    \displaystyle C&=& \alpha_A \nonumber \\ 
    \displaystyle \eta&=&\frac{B - \sqrt{B^2 - 4AC}}{2A}, 
\end{eqnarray}
we obtain the neutral fraction $\eta$ for a given density $n(\mathrm{H})$, temperature $T$ and photo-ionisation rate. With $\eta$, we calculate the total neutral mass associated with the neutral phase as $m_{\mathrm{neutral}}= \eta \times  m_{\mathrm{SPH}} $ and at the same time define the total ionised gas mass as $m_{\mathrm{ionised}}= m_{\mathrm{SPH}} -  m_{\mathrm{neutral}}$. 

We take the background interstellar radiation field (ISRF) over the SPH particle due to star formation into account as another important parameter. For this, we assume that gas in the disk is self-gravitating and obeys the Kennicutt-Schmidt (KS) star-formation law \citep{2008MNRAS.383.1210S}. We calculate the Jeans length as
\begin{eqnarray}
     \displaystyle L_J=\frac{c_s}{\sqrt{G \rho}} \, \, \mathrm{with} \label{eq:JeansL}  \\
     \displaystyle c_s = \sqrt{\frac{\gamma  P_{\mathrm{Tot}}}{\rho}},
\end{eqnarray}
where $c_s$ is the effective sound speed \citep{2008MNRAS.383.1210S,2001ApJ...559..507S}, and $\gamma=5/3$ is the adiabatic index (different from the polytropic index $\gamma_{\mathrm{lim}}$). 
The total pressure for the SPH particle \citep[entropy-weighted,][]{2017arXiv170609899T} is defined as
\begin{equation}
    \displaystyle P_{\mathrm{Tot}} = E \rho^{\gamma},
    \label{eq:Ptot}
\end{equation}
where $E$ is the entropy. 
Then the SFR density is 
\begin{eqnarray}
    \rho_{\mathrm{SFR}}=A \rho \left(1 \mathrm{M_{\sun}}/\mathrm{pc}^2  \right)^{-n} \left( \frac{\gamma}{G} f_g P_{\mathrm{Tot}} \right)^{(n-1)/2},
    \label{eq:SFRden}
\end{eqnarray}
where the KS law exponent is $n=1.4$, $G$ is the gravitational constant, the gas fraction $f_g$ is assumed to be unity and the absolute star-formation efficiency is $A=1.515 \times 10^{-4} \, \mathrm{M_{\sun}} \, \mathrm{yr}^{-1} \, \mathrm{kpc}^{-2}$ \citep{2008MNRAS.383.1210S,2017arXiv170609899T}. The SFR surface density is defined from Eqs.~\ref{eq:JeansL} and \ref{eq:SFRden} as $\Sigma_{\mathrm{SFR}} = \rho_{\mathrm{SFR}} L_{J}$. Following \citet{2015MNRAS.452.3815L}, we determine the background ISRF in units of the Habing radiation field \citep[$ G_0=1.6 \times 10^{-3} \, \mathrm{erg} \, \mathrm{cm}^{-2} \, \mathrm{s}^{-1} $]{1968BAN....19..421H} as 
\begin{eqnarray}
    G_{\mathrm{0,bkg}}^{'}=\frac{\Sigma_{\mathrm{SFR}}}{\Sigma_{\mathrm{SFR,MW}}},
\end{eqnarray}
where $\Sigma_{\mathrm{SFR,MW}}=0.001 \mathrm{M_{\sun}} \, \mathrm{yr}^{-1} \, \mathrm{kpc}^{-2}$ is the value of the SFR surface density in the solar neighbourhood \citep{2011MNRAS.415.2827B}. 

We calculated the ISRF coming directly from the local stars close to the gas particles (inside the smoothing length $h$ of the gas particle) following the procedure described by \citet{2017ApJ...846..105O}. We used the information from \verb|starburst99|  \citep{2014ApJS..212...14L} stellar models with a mass of $10^4 \mathrm{M_{\sun}}$ to obtain a grid of models at different metallicities\footnote{We adopt the Geneva stellar models \citep{1992A&AS...96..269S} with standard mass loss.} and ages\footnote{The star formation time $t_*$ is used to calculate the age of the stars (i.e. lookback time).} for the star particles. A Kroupa initial mass function (IMF) was adopted from \verb|starburst99| although the \textsc{EAGLE} simulations used a different IMF (Chabrier); the expected differences in the FUV luminosities ($L_{\mathrm{FUV}}$) are negligible ($<2\%$) in these range of parameters \citep{2017ApJ...846..105O}. From these stellar models, we obtained the respective $L_{\mathrm{FUV}}$ (calculated as the average luminosity between 912 and 2066 \AA, $6-13.6$ eV) for each metallicity and age. Then we interpolated these values to estimate the $L_{\mathrm{FUV}}$ for each of the star particles in the galaxy.

The local radiation field is then defined as
\begin{eqnarray}
    & \displaystyle\frac{G_{\mathrm{0,loc}}}{\mathrm{erg}\, \mathrm{cm}^{-2} \mathrm{s}^{-1}}& \nonumber \\
    &= \displaystyle\sum_{|r_{\mathrm{gas}}- r_{*,i}|<h}
    \frac{L_{\mathrm{FUV,i}}}{4 \pi |r_{\mathrm{gas}}- r_{*,i}|^{2}} \frac{m_{*,i}}{10^{4} \mathrm{M_{\sun}}},
\end{eqnarray}
where $|r_{\mathrm{gas}}- r_{*,i}|$ is the distance between the gas and star particles, $m_{*,i}$ is the stellar mass and $L_{\mathrm{FUV,i}}$ the value of the FUV luminosity for each star particle. We then define the total ISRF hitting the outer part of the neutral cloud as 
\begin{equation}
    G_{\mathrm{0,cloud}}=G_{\mathrm{0,bkg}}+G_{\mathrm{0,loc}},
\end{equation}
where we have normalised the value of $G_{\mathrm{0,bkg}}^{'}$ by $G_{\mathrm{0,MW}}$ (i.e. $G_{\mathrm{0,bkg}}^{'} =G_{\mathrm{0,bkg}}/G_{\mathrm{0,MW}}$), using $G_{\mathrm{0,MW}}=9.6 \times 10^{-4}$ erg cm$^{-2}$ s$^{-1}$ \citep{2011ApJS..196...15S}. Typical values for $G_{\mathrm{0,cloud}}$ are in the range $\sim 0.1$--$10^8$. The low ranges ($\sim 0.1$) are regions where the gas particles are only affected by $G_{\mathrm{0,bkg}}$, while at high ranges ($\sim 10^8$) $G_{\mathrm{0,loc}}$ is dominant as the stars particles are very close to the gas particles.   

\subsubsection{Neutral clouds}\label{subsec:PDRMCModel}

Following the work of \citet{2015ApJ...814...76O}, we identify two phases in the environments of the neutral gas clouds that we analyse in this work, the dense molecular gas and the neutral atomic gas. To determine the properties of these phases, the neutral mass $m_{\mathrm{neutral}}$ is divided into the neutral clouds by sampling the mass function as observed in the Galactic disk and Local Group:
\begin{equation}
    \frac{dN}{dm_{\mathrm{cloud}}} \propto m_{\mathrm{cloud}}^{-\beta},
    \label{eq:GMClaw}
\end{equation}
with $\beta=1.8$ \citep{2007prpl.conf...81B}. We apply a lower cut on the mass of $10^4\,\mathrm{M_{\sun}}$ and an upper cut of $10^6\,\mathrm{M_{\sun}}$ \citep{2008ApJS..176..331N,2008ApJ...684..996N}. The remaining gas below the lower limit of $10^4\,\mathrm{M_{\sun}}$ is discarded and so it goes to the ionised mass ($m_{\mathrm{ionised}}$); in general this is below 1\% of the mass. The neutral clouds are randomly distributed within 0.5$h$, with the radial displacement scaling inversely with $m_{\mathrm{cloud}}$, to conserve the mass distribution in the galaxy. Changing this radial distribution limit ($0.5h$) affects the final luminosities minimally. 

We use the entropy to obtain the external pressure (Eq.~\ref{eq:Ptot}) and to calculate the radius of the neutral cloud ($R_{\mathrm{cloud}}$). We take the relative contributions by the cosmic-ray (CR) and magnetic pressure into account, where $\alpha_0 = 0.4$ and $\beta_0 = 0.25$ following \citet{1989ApJ...344..306E}:
\begin{equation}
    P_{\mathrm{ext}}= \frac{P_{\mathrm{Tot}} }{1 + \alpha_0 + \beta_0}.
\end{equation}
$R_{\mathrm{cloud}}$ is then obtained following the virial theorem from a pressure normalised mass-size relation as \citep{2011MNRAS.416..710F,2013ApJ...779...46H,2018ApJ...857...19F}
\begin{equation}
    \frac{R_{\mathrm{cloud}}}{\mathrm{pc}}=\left( \frac{P_{\mathrm{ext}}/k_B}{10^4 \mathrm{cm}^{-3} \mathrm{K}} \right)^{-1/4} \left( \frac{m_{\mathrm{cloud}}}{290 \mathrm{M_{\sun}}}\right)^{1/2},
    \label{eq:GMCMassSize}
\end{equation}
resulting in sizes of $R_{\mathrm{cloud}}\approx 1$--$300\,\mathrm{pc}$ (Sect.~\ref{subsec:ISMSum}). For the density inside the neutral cloud, we use a gas distribution described by a Plummer profile:
\begin{equation}
    n(\mathrm{H})(R) = \frac{3m_{\mathrm{cloud}}}{4\pi R_p^{3}} \left( 1+ \frac{R^2}{R_p^2} \right)^{-5/2}, \label{eq:Plummer}
\end{equation}
with $R_p = 0.1 R_{\mathrm{cloud}}$. Adopting this density profile ensures a finite
central density. In addition, \citet{{2019MNRAS.482.4906P}} showed that this profile is better at reproducing the {[\ion{C}{II}]} luminosity in galaxies from $z=0$ to $z=6$ compared to other distributions (power law, logotropic and constant density profiles). With these values, we describe the structure for the neutral clouds. In addition, the neutral clouds inherit some physical parameters (e.g. $Z$, $G_{0,\mathrm{bkg}}^{'}$, $X_C$ among others) from the SPH particle.

The {[\ion{C}{II}]} emission arises from within the neutral clouds except from the inner core region, which is shielded from FUV radiation. We calculate the radius where the abundances of C and $\mathrm{C}^+$ are equal ($R_{\mathrm{CI}}$) and assume that all the atoms inside this radius are neutral, so the emission of {[\ion{C}{II}]} in that region is zero. We follow the steps described in \citet{2015ApJ...814...76O}, who used the approach of \citet{2006A&A...451..917R} with the following reactions for the formation and destruction of $\mathrm{C}^+$:
\begin{eqnarray}
    \mathrm{C} + \gamma &\longrightarrow& \mathrm{C}^{+} + e^{-} \label{eq:form}\\
    \mathrm{C}^{+} + e^{-} &\longrightarrow& \mathrm{C} + \gamma \label{eq:destruc}\\
    \mathrm{C}^{+} + \rm{H}_{2} &\longrightarrow& \mathrm{CH}_{2}^{+} + \gamma.  \label{eq:destruc2}
\end{eqnarray}
We solve $R_{\mathrm{CI}}$ with the equation
\begin{eqnarray}
     \displaystyle 5.13 \times 10^{-10} s^{-1} \chi G_{\mathrm{0,cloud}}  = n(\mathrm{H})(R_{\mathrm{CI}}) \left[ a_C X_C + 0.5 k_C \right] \label{eq:RCI}
\end{eqnarray}
with 
\begin{eqnarray}
    \displaystyle \chi = \int_1^{\infty} \frac{\exp{(-\mu \xi_{\mathrm{FUV}} A_V(R_{\mathrm{CI}}))}}{\mu^{2}} d\mu,
\end{eqnarray}
where the left-hand side of Eq.~\ref{eq:RCI} is C$^{+}$ formation (Eq.~\ref{eq:form}) and the right-hand side is  C$^{+}$ destruction (Eqs.~\ref{eq:destruc} and \ref{eq:destruc2}). We use the same constants as \citet{2015ApJ...814...76O} for the recombination and radiative association rate coefficients: $a_C= 3 \times 10^{-11} \mathrm{cm}^{3} \mathrm{s}^{-1}$ and $k_C=8 \times 10^{-16} \mathrm{cm}^{3} \mathrm{s}^{-1}$ \citep{2009ApJ...707..954P}. The normalisation constant ($5.13 \times 10^{-10} s^{-1}$) comes from the ionisation rate in the photo-dissociation region (PDR) model by \citet{2006A&A...451..917R}. In the integral, the isotropic radiation field is accounted by $\mu= 1/\cos{\theta}$, where $\theta$ is the angle between the Poynting vector and the normal direction. The visual extinction correspondingly is defined as $A_V (R_{\mathrm{CI}}) = 0.724 \sigma_{\mathrm{dust}} Z^{'} \left< n(\mathrm{H}) \right> R_{\mathrm{cloud}} \ln{\left(R_{\mathrm{cloud}}/R_{\mathrm{CI}}\right)}$ \citep{2009ApJ...707..954P}, where $\sigma_{\mathrm{dust}} = 4.9 \times 10^{-22} \mathrm{cm}^{2}$ is the FUV dust absorption cross section \citep{1982A&A...105..372M}, $Z^{'}$ is the mean metallicity of the galaxy and $\left< n(\mathrm{H}) \right>$ is the average density inside the neutral cloud. The difference in the opacity between visual and FUV light is set to $\xi_{FUV}=3.02$ \citep{2006A&A...451..917R}. The carbon abundance relative to hydrogen $X_C$ comes from the carbon mass fraction of the parent SPH particle. For simplicity, \citet{2006A&A...451..917R}  assumed that the density of electrons is similar to that of carbon to obtain Eq.~\ref{eq:RCI}. 

As these calculations are computationally expensive, we create a grid of four variables to solve for $R_{\mathrm{CI}}$, using the following ranges: $4 \leq \log(m_{\mathrm{cloud}} [M_\sun]) \leq6$, $18.5 \leq \log(R_{\mathrm{cloud}} [cm] ) \leq 21$,  $-1.5 \leq \log(G_{\mathrm{0,cloud}}) \leq 8$ and $-7 \leq \log(X_C) \leq -1.5$, all in steps of 0.125 dex. With this grid of $\sim$1.2 million points, we obtain a solution for $R_{\mathrm{CI}}$ for each neutral cloud. 

To differentiate neutral atomic gas and dense molecular gas, we need to define a radius at which molecular hydrogen dominates. We calculate the molecular H$_2$ fraction following \citet{2011ApJ...728...88G} as
\begin{equation}
    \displaystyle f_{\mathrm{H}_2}= \left( 1 + \frac{\tilde{\Sigma}}{\Sigma_{HI + H_2}} \right)^{-2},
\end{equation}
where 
\begin{eqnarray}
    \tilde{\Sigma} &=& 20 \mathrm{M_{\sun}} pc^{-2} \times \nonumber\\
    &&\frac{\Lambda^{4/7}}{D_{\mathrm{MW}}} \frac{1}{\sqrt{1 + U_{\mathrm{MW}} D_{\mathrm{MW}}^2}},\\
    \displaystyle \Lambda &=& \ln{ \left[ 1 + g D_{\mathrm{MW}}^{3/7} \left( U_{\mathrm{MW}}/15 \right)^{4/7} \right]},\\
    g &=& \frac{1 + \alpha s + s^2}{1 +s},\\
    s &=& \frac{0.04}{D_* + D_{\mathrm{MW}}},\\
    \alpha &=& 5 \frac{U_{\mathrm{MW}}/2}{1 + \left( U_{\mathrm{MW}}/2 \right)^{2} }, \\
    D_* &=& 1.5 \times 10^{-3} \ln{ \left[ 1 + \left( 3 U_{\mathrm{MW}} \right)^{1.7} \right] },
\end{eqnarray}
the metallicity of gas in solar units is $D_{\mathrm{MW}}= Z_{\mathrm{gas}} / Z_{\sun}$, and the local UV background is $U_{\mathrm{MW}}= \mathrm{SFR}/\mathrm{SFR}_{\mathrm{MW}}$. Then the molecular fraction can be used for determining the radius at which the transition from atomic to molecular H occurs ($R_{\mathrm{H}_2}$) in a Plummer profile (Eq.~\ref{eq:Plummer}):
\begin{equation}
    f_{\mathrm{H}_2}= \left( \frac{R_{\mathrm{H}_2}}{R_{\mathrm{cloud}}} \right)^{3} \left( \frac{R_\mathrm{p}^2 + R_{\mathrm{cloud}}^2}{R_\mathrm{p}^2 + R_{\mathrm{H}_2}^2} \right)^{3/2}. 
    \label{eq:fh2}
\end{equation}

This information helps us to calculate the density and temperature at $R_{\mathrm{H}_2}$ in the dense molecular gas and neutral atomic gas regions.

\subsubsection{Diffuse ionised gas}\label{subsec:DIGModel}

For the diffuse ionised gas (DIG), we follow a slightly different approach than that used by \citet{2015ApJ...814...76O,2017ApJ...846..105O}, where they assumed a DIG cloud with a radius equal to the smoothing length ($h$). We want to avoid dependency on the simulation's resolution and overproduction of DIG in SPH particles with very large $h$ ($\gtrsim$ 5 kpc in 100 Mpc boxes, Sect.~\ref{subsec:ISMSum}). We instead calibrate our models with the estimated luminosity of {[\ion{N}{II}]} lines at 122 and 205 $\mu$m emitted almost entirely from the ionised medium. By comparing these lines it is possible to deduce the fractions of the ionised gas \citep{2017ApJ...845...96C}. For example {[\ion{N}{II}]} at 122 $\mu$m is associated with the DIG rather than with a compact \ion{H}{II} region \citep{2012A&A...548A..20C}.

To calibrate the DIG fraction in the {[\ion{C}{II}]} line, we create a distribution of radii of the DIG ($R_{\mathrm{DIG}}$) assuming an isothermal sphere with uniform density. We assume that the distribution function behaves as a smoothed broken power law for $R_{\mathrm{DIG}}$:

\begin{equation}
    p(R_{\mathrm{DIG}})= \left( \frac{R_{\mathrm{DIG}}}{R_b} \right)^{-\alpha_1} \left\{ \frac{1}{2} \left[ 1+ \left( \frac{R_{\mathrm{DIG}}}{R_b} \right)^{1/\Delta} \right] \right\}^{( \alpha_1 - \alpha_2 ) \Delta},
    \label{eq:RDIG}
\end{equation}
where $R_{b}$ is the break radius, $\alpha_1$ is the power law index for $R_{\mathrm{DIG}} \ll R_b$, $\alpha_2$ is the power law index for $R_{\mathrm{DIG}} \gg R_b$ and $\Delta$ is the smoothness parameter. We create a grid of different values as described in Table~\ref{tab:GridValuesCalibration} for the parameters in Eq.~\ref{eq:RDIG}. We fix the $\Delta$ parameter to 0.1 to achieve a distribution with a smooth peak. We estimate the line luminosity for the given grid of values in a random sample (20\%) of galaxies from \textsc{Ref-L0100N1504}. We only use \textsc{Ref-L0100N1504} for the DIG calibration as the small simulation boxes do not contain enough galaxies to compare with the observational sample.

\begin{table}
    \caption{Grids for the calibration of Eq.~\ref{eq:RDIG} used in this work. The resulting total number of grid points is 4131.}
    \label{tab:GridValuesCalibration}
    \centering
    \begin{tabular}{lcrrr}
    \hline
    \hline
    Parameter & Unit & Min. & Max. & Interval\\
    \hline
    log $R_b$ &[kpc] &$-0.8$&0.5&0.026\\
    $\alpha_1$ & -- &$-5.0$&$-1.0$&0.5\phantom{00}\\
    $\alpha_2$ & -- &1.0&5.0&0.5\phantom{00}\\
    \hline
    \end{tabular}    
\end{table}

We use observational data from \citet{2016ApJS..226...19F} to calibrate the contribution of the DIG. We use the $L_{\mathrm[\ion{N}{II}]122}$--SFR, $L_{\mathrm[\ion{N}{II}]205}$--SFR and $L_{\mathrm[\ion{N}{II}]122}/L_{\mathrm[\ion{N}{II}]205}$--SFR relations to compare the observational and simulated samples. We consider only cases with a SFR inside the values that the selected simulation could achieve ($-2.8<\log(\mathrm{SFR})<1.7$), for a total of 69 observed galaxies. The two datasets (observational and simulated) are binned by SFR into identical bins. We calculate the mean and standard deviation in each bin to compare them using a chi-square ($\chi^2$) test. We notice that $\chi^2$ values were not significantly affected by the $\alpha$ parameters, so we selected $\alpha_1 = -2.0$ and $\alpha_2$ = 1.0, assuming that the DIG size is larger than the neutral clouds. We found that the $R_b$ value that minimises the $\chi^2$ is $R_b \approx$ 900 pc.

\begin{figure*}[ht]
 \includegraphics[width=\textwidth]{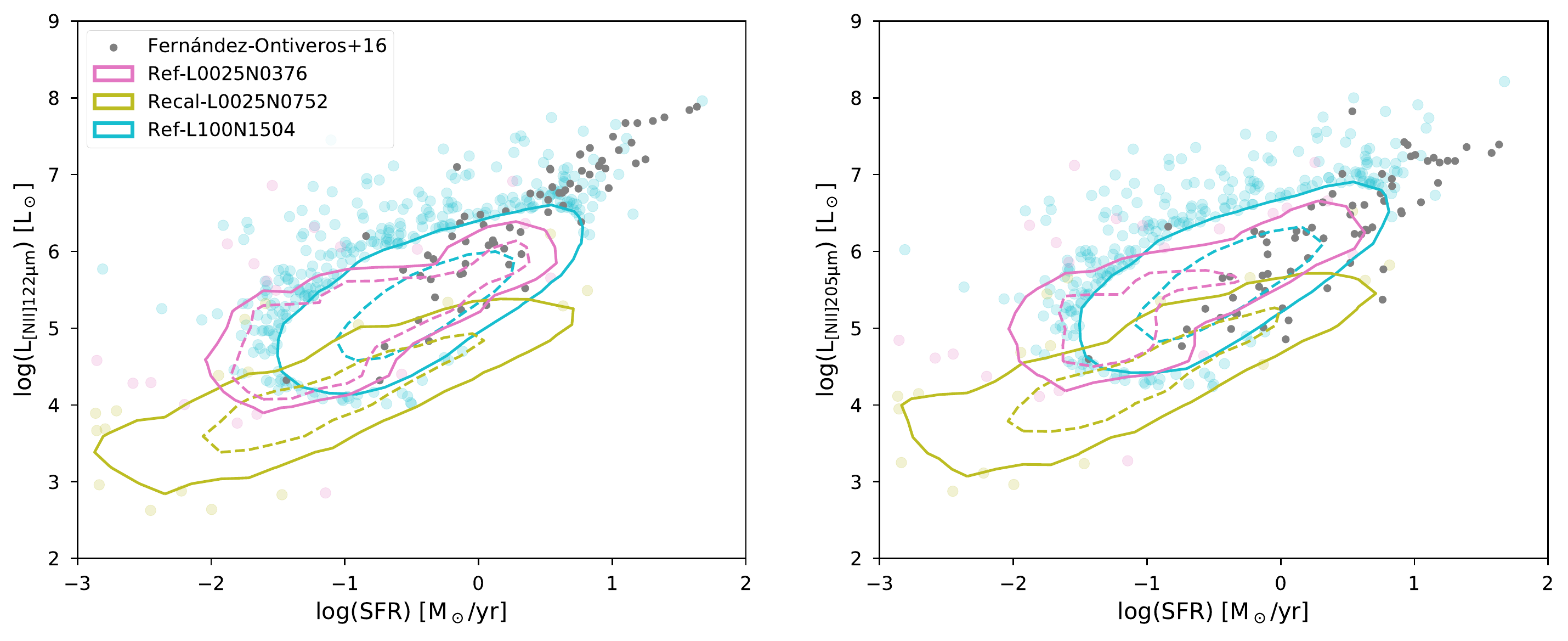}
\caption{$L_{\mathrm[\ion{N}{II}]122}$--SFR (left) and $L_{\mathrm[\ion{N}{II}]205}$--SFR (right) relations for the three simulations used (\textsc{Ref-L0025N0376}, \textsc{Recal-L0025N0752}, and \textsc{Ref-L0100N1504}) and the observational sample used for the DIG calibration from \citet{2016ApJS..226...19F} (grey dots). The smoothed contours show where 90\% (solid) and 50\% (dashed) of the galaxies from the respective simulations lie. The remaining 10\% of galaxies are represented as dots using the same colours.}
 \label{fig:z0NII}
\end{figure*}

In Fig.~\ref{fig:z0NII} we present the $L_{\mathrm[\ion{N}{II}]}$--SFR relations at $z=0$ from the simulations used in this work compared with the sample of galaxies with {[\ion{N}{II}]} used in the DIG calibration from \citet{2016ApJS..226...19F}. We recover the same trend of the observed {[\ion{N}{II}]} line luminosities with \textsc{Ref-L0100N1504}. The DIG calibration shows a good agreement for simulations with the same resolutions (\textsc{Ref-L0100N1504} and \textsc{Ref-L0025N0376}) but \textsc{Recal-L0025N0752} is off by less than 1 dex, which is related to the re-calibration (see discussion in Sect.~\ref{subsec:DisRightSim}).

We randomly sample the DIG radii using Eq.~(\ref{eq:RDIG}) for each galaxy. We use the assigned DIG radii and the ionised mass ($m_{\mathrm{ionised}}$) to calculate the density of this ISM phase for each SPH particle. These lead us to densities ranging from around $10^{-6}$ to 3 $\mathrm{cm}^{-3}$ for the DIG (due to the limits in \textsc{Cloudy} cooling tables, Table~\ref{tab:GridValues}), with a peak at $10^{-2} \mathrm{cm}^{-3}$ (see results in Sect.~\ref{subsec:ISMSum}).

\subsection{Line luminosity prediction} \label{subsec:CloudyGrid}

\begin{figure*}[ht]
 \includegraphics[width=\textwidth]{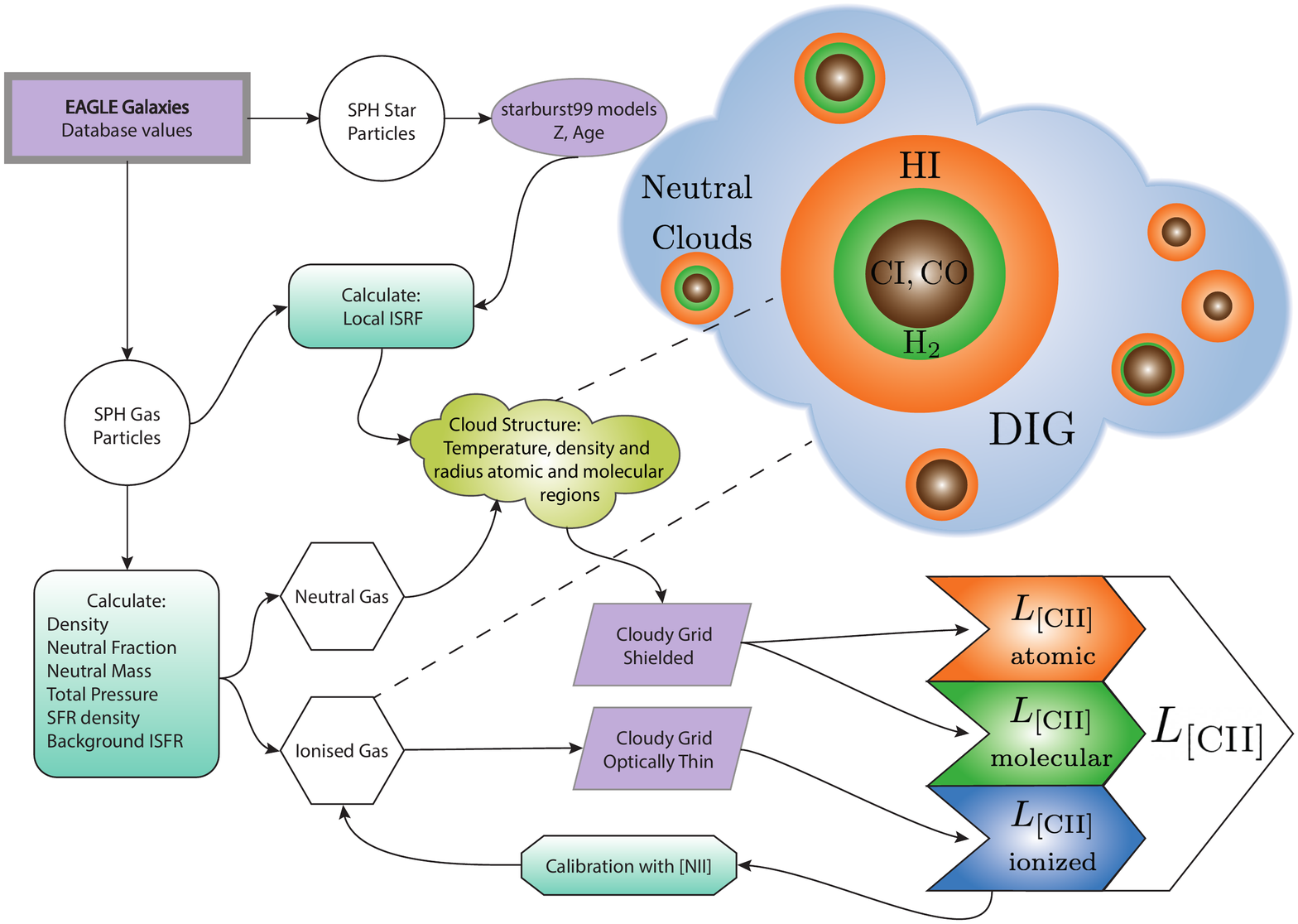}
 \caption{Flowchart of the sub-grid procedures applied to the SPH simulation to simulate line emission in post-processing. For each \textsc{EAGLE} simulation, we obtain the galaxy information from the database and we use the star and gas particle data to implement the ISM model. Next, we calculate physical properties in each phase to obtain the neutral clouds and DIG structures. We use \textsc{Cloudy} cooling tables \citep{2020MNRAS.497.4857P} to get the line luminosity for {[\ion{C}{II}]}. We first calculate $L_{\mathrm{[\ion{C}{II}],DIG}}$ and then calibrate the ionised gas using the predicted luminosities of the {[\ion{N}{II}]} lines. We then obtain the total line luminosity from the luminosities of the different ISM phases. The dashed lines connect the gas environment to the ingredients of the model.}
 \label{fig:ISMIlustration}
\end{figure*}

We predict line luminosities with the help of \textsc{Cloudy} v17.01 \citep{2017RMxAA..53..385F}, using a set of cooling tables presented by \citet{2020MNRAS.497.4857P}. \textsc{Cloudy} is a 1D spectral synthesis code that predicts atomic and molecular line intensities in different environments using an escape probability formalism. Here we describe briefly the important aspects of the cooling tables used. 

An equally spaced grid is used in the dimensions redshift $z$, gas temperature $\log T$, metallicity $\log Z$, and gas density $\log n(\mathrm{H})$, as presented in Table~\ref{tab:GridValues}. In this work, we only made use of the redshift bin at $z=0$. The importance of using the redshift as a parameter resides on the non-negligible effect on the line emissivity from the cosmic microwave background (CMB). The spin temperature of the {[\ion{C}{II}]} emission is similar to the CMB temperature at high redshift in low density environments (WNM), affecting the {[\ion{C}{II}]} emission. In contrast, at high densities (CNM) the spin temperatures are larger, so the {[\ion{C}{II}]} emission is not affected \citep{2015ApJ...813...36V,2017MNRAS.471.4128P,2018A&A...609A.130L,2018ApJ...857..148O}. However, for this study at $z=0$, the CMB is not important.

\textsc{Cloudy} is used to propagate the incident radiation in a plane-parallel gas until the column density $N_{\mathrm{sh}}$ is reached. For gas with temperatures of $T \le 10^3\,\mathrm{K}$ the shielding column $N_{\mathrm{sh}}$ is assumed to be equal to one half of the Jeans column density
\begin{equation}
    \log N_J [\mathrm{cm}^{-2}] = 19.44 + 0.5 \times \left( \log n(\mathrm{H}) [\mathrm{cm}^{-3}] + \log T [\mathrm{K}]  \right)
\end{equation}
to model the shielding from the edge to the centre of a self-gravitating gas cloud with an extent equal to the Jeans length $\lambda_{\mathrm{J}}$.
For higher temperatures, the column density of the self-shielding gas asymptotically approaches that of the optically thin gas and a maximum column density of $N_{\mathrm{max}} =10^{24}\,\mathrm{cm}^{-2}$ and a maximum length scale of $l_{\mathrm{max}} = 100\,\mathrm{kpc}$ are imposed.

For the radiation field, redshift-dependent contributions from the CMB and UV background \citep{2020MNRAS.493.1614F} are applied. In addition, the ISRF (`table ism' in \textsc{Cloudy}) and cosmic rays (CR), scaled to solar neighbourhood values, are added depending on the column density. Solar abundances are assumed, with the abundance ratios modified by dust depletion following \citet{ 2009ApJ...700.1299J}. Primordial abundances are also calculated (when $\log$ $Z/Z_{\sun} = -50$) using \citet{2016A&A...594A..13P} primordial values for helium (not used in this work). 

The `Orion' grain distribution (from \textsc{Cloudy}) is used to take into account other physical effects from dust (e.g. photoelectric heating and charge and collisional processes) by assuming a dust-to-gas ratio dependent on the metallicity and column density \citep[assuming $\log{N_{\mathrm{H}} [\mathrm{cm}^{-2}]} = 20.56$ from the gas surface density in the solar neighbourhood,][]{2015MNRAS.452.3815L}. Furthermore, polycyclic aromatic hydrocarbons (PAHs) are included but quantum heating is disabled (no `qheat'). The large (i.e. more detailed) H2 model in Cloudy is used and the CR photo-dissociation rate is re-scaled to match the UMIST database\footnote{\url{udfa.net}} values \citep[see also][]{2020RNAAS...4...78S}.

The two cooling tables we use in this work, ideal for cosmological simulations, have the ISRF and CR rate values reduced by 1 dex relative to the MW values \citep{1987ASSL..134..731B,2007ApJ...671.1736I} to better match the observations of the transition between atomic and molecular hydrogen \citep{2020MNRAS.497.4857P}. In other words, physical parameters decrease: the SFR surface density $\Sigma_{\mathrm{SFR}}$ decreases from $10^{-3} \, \mathrm{M_{\sun}} \, \mathrm{yr}^{-1} \, \mathrm{kpc}^{-2}$ to $10^{-4} \, \mathrm{M_{\sun}} \, \mathrm{yr}^{-1} \, \mathrm{kpc}^{-2}$, and the CR hydrogen ionisation rate $\log \zeta$ decreases from $-15.7\,\mathrm{s^{-1}}$ to $-16.7\,\mathrm{s^{-1}}$. The first table includes self-shielding and the second table is the optically thin counterpart (1-zone unattenuated cloud) of the first table\footnote{The optically thick and optically thin tables are the UVB\_dust1\_CR1\_G1\_shield1 and UVB\_dust1\_CR1\_G1\_shield0 in \citet{2020MNRAS.497.4857P}, respectively.}. The latter provides us with the emissivities of the DIG, approximating this phase as optically thin. We sample and limit the grids to the values presented in Table~\ref{tab:GridValues}. 

We use thermal equilibrium temperatures, where cooling is equal to heating, which depend mainly on two variables $n(\mathrm{H})$ and $Z$. The temperatures of the gas phases are then obtained with a linear interpolation of the grid with density and metallicity (as other parameters like ISRF or CR depend on these) for the neutral atomic gas, dense molecular gas and DIG phases. This assumption is probably incorrect for the DIG, as ionised regions could be outside thermal equilibrium, because DIG gas can have higher temperatures, but temperature values $\approx 10^4\,mathrm{K}$ are commonly assumed when modelling ionised regions \citep[e.g.][]{2009RvMP...81..969H,2019MNRAS.488.1977V,2019MNRAS.489....1F,2020MNRAS.497.4857P}. For atomic and molecular gas, thermal equilibrium is commonly used in PDR models \citep{1985ApJ...291..722T, 1999RvMP...71..173H}. 

\begin{table}
    \caption{Sampling of gas properties in the \textsc{Cloudy} grid used in this work. The resulting number of grid points is 610 for the shielded and optically thin calculations.}
    \label{tab:GridValues}
    \centering
    \begin{tabular}{lcccc}
    \hline
    \hline
    Parameter & Unit & Min. & Max. & Interval\\
    \hline
    log $Z$ &[$Z_{\sun}$] &$-$4.0&0.5&0.5\\
    log $n(\mathrm{H})$ &[cm$^{-3}$] &$-$6.0&6.0&0.2\\
    \hline
    \end{tabular}    
\end{table}

The {[\ion{C}{II}]} line luminosity is then computed from the integral of the {[\ion{C}{II}]} emissivity, $\Lambda_{\mathrm{[\ion{C}{II}]}}$, interpolated from the \textsc{Cloudy} cooling table as described above, through the equation

\begin{equation}
    L_{\mathrm[\ion{C}{II}]} = \int \Lambda_{\mathrm{[\ion{C}{II}]}} dV = 4 \pi \int_{R_1}^{R_2} \Lambda_{\mathrm{[\ion{C}{II}]}} R^2 dR.
\end{equation}

\noindent The limits of the integral are defined by the three components of the ISM model, DIG, atomic gas and molecular gas; therefore, $R_1 = 0$ and $R_2 = R_{\mathrm{DIG}}$ for the DIG, $R_1 = R_{\mathrm{H}_2}$ and $R_2 = R_{\mathrm{cloud}}$ for the atomic gas and $R_1 = R_{\mathrm{CI}}$ and $R_2 = R_{\mathrm{H}_2}$ for the molecular gas. Ten shells are used to estimate the luminosities for the atomic gas and molecular gas where we assume that the density follows a Plummer profile (Eq.~\ref{eq:Plummer}). In the case of the DIG, one shell is assumed because we assume a homogeneous density distribution. 

\subsection{Summary and verification} \label{subsec:ISMSum}

To summarise, we illustrate the path from the \textsc{EAGLE} simulations (Sect.~\ref{subsec:EAGLEDes}) to the total luminosity of {[\ion{C}{II}]} (Sect.~\ref{subsec:CloudyGrid}) in Fig.~\ref{fig:ISMIlustration}. First, we select a sample of galaxies from the EAGLE database and retrieve the gas and star particle data of the sample. Second, we calculate physical properties such as density and neutral fraction for the gas particles. Local ISRF is calculated with \texttt{starburst99} models derived from star particles, while background ISFR comes from the SFR surface density. The neutral gas mass and the ISRF are used to create the neutral cloud structures inside galaxies, while the ionised gas mass is used to create the DIG. After calibration with the {[\ion{N}{II}]} lines for the DIG, we obtain the luminosity of {[\ion{C}{II}]} line for each phase using \textsc{Cloudy} cooling tables \citep{2020MNRAS.497.4857P}. Finally, we calculate the total {[\ion{C}{II}]} luminosity in a galaxy with the contributions from the three ISM phases combined.

\begin{figure}[ht]
 \includegraphics[width=\columnwidth]{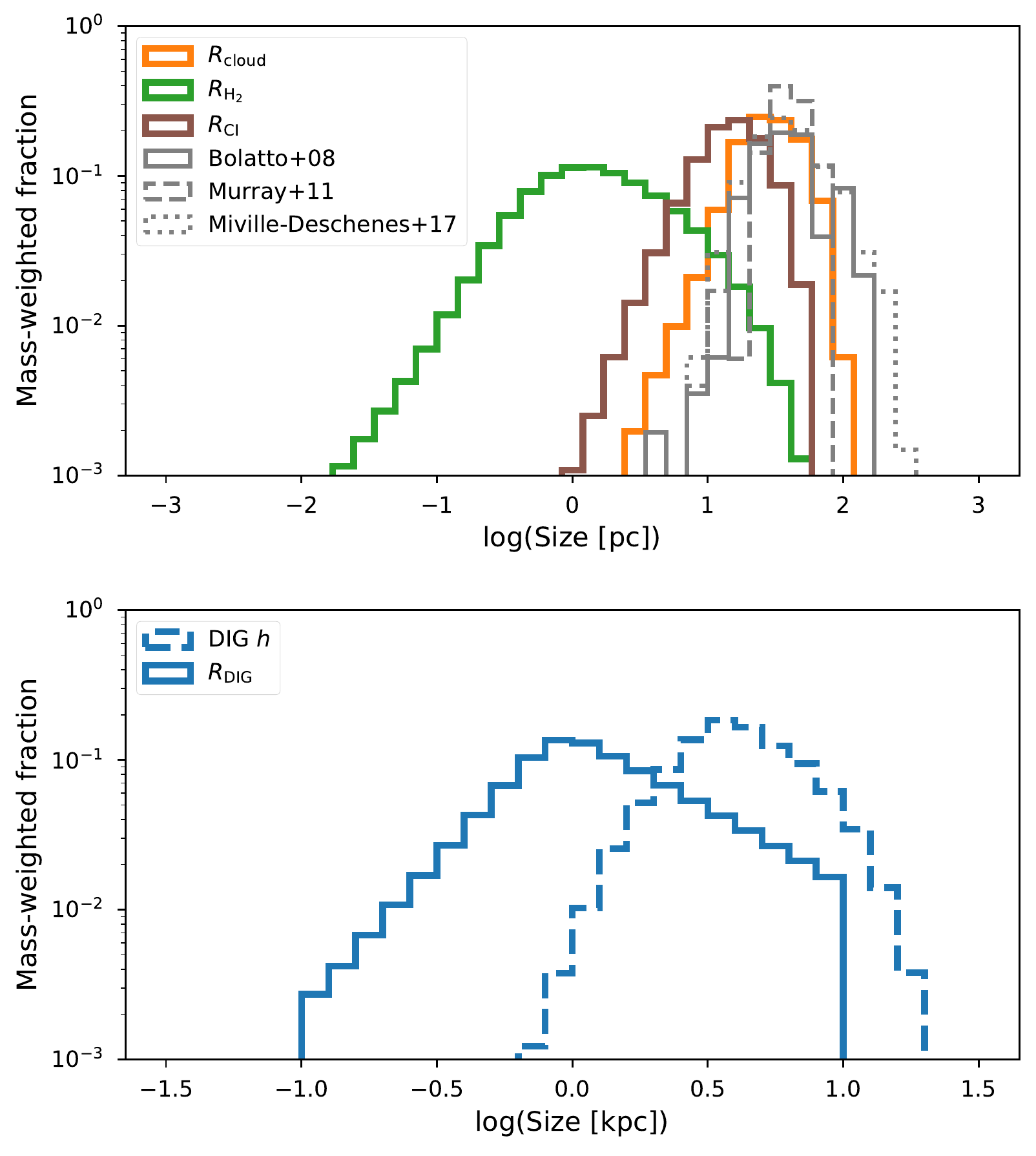}
 \caption{Mean mass-weighted distribution of the sizes of the ISM phases for \textsc{Recal-L0025N0752}. \textit{Upper panel}: Radii defining the limits of the phases inside the neutral clouds: $R_{\mathrm{cloud}}$, $R_{\mathrm{H}_2}$ and $R_{\mathrm{CI}}$. We compare with sizes of MW clouds \citep[solid and dashed grey lines,][]{2011ApJ...729..133M,2017ApJ...834...57M} and local Universe clouds \citep[dotted grey lines]{2008ApJ...686..948B}.
 \textit{Lower Panel}: Assumed mean mass-weighted distribution of the radii of the DIG using the smoothing length $h$ (dashed) or Eq.~\ref{eq:RDIG} (solid) to define $R_{\mathrm{DIG}}$.}
 \label{fig:Sizes}
\end{figure}

We plot the distribution of the sizes for the neutral clouds and DIG (described in  Sect.~\ref{subsec:ISMModel}) in Fig.~\ref{fig:Sizes}. The values for $R_{\mathrm{H}_2}$ show that the molecular regions are in general smaller than $R_{\mathrm{CI}}$. As the neutral clouds are modelled with a Plummer profile (Eq.~\ref{eq:fh2}), the molecular fraction will be restricted to the centre of the neutral clouds at around one pc, so the emission coming from these regions will be less than the atomic gas emission in most cases. In cases where $R_{\mathrm{H}_2}$ is  very small ($f_{\mathrm{H}_2}$ near zero and $R_{\mathrm{CI}} > R_{\mathrm{H}_2}$), the atomic phase will dominate between $R_{\mathrm{cloud}}$ and $R_{\mathrm{CI}}$. For the case when the shielding is efficient ($f_{\mathrm{H}_2} \gg 0$), the atomic gas goes from $R_{\mathrm{cloud}}$ to $R_{\mathrm{H}_2}$ and the molecular goes from $R_{\mathrm{H}_2}$ to $R_{\mathrm{CI}}$. Therefore, $R_{\mathrm{CI}}$ is the dominant internal bound for the atomic region in most cases. We sketch these radii in the different neutral clouds in Fig.~\ref{fig:ISMIlustration}.

Typical sizes for these neutral clouds ($\sim 0.5 - 200$ pc) agree with observed values in the MW \citep[e.g.][]{2011ApJ...729..133M,2017ApJ...834...57M} and in the local Universe  \citep[e.g.][]{2008ApJ...686..948B}, as presented in the upper panel of Fig.~\ref{fig:Sizes}. In the case of the DIG, we compare two different cases, using the smoothing length $h$ or Eq.~\ref{eq:RDIG}. When using $h$, the DIG sizes can go up to $\sim20$ kpc, which is the size of some galaxies in the simulations, while with $R_{\mathrm{DIG}}$ from Eq.~\ref{eq:RDIG}, sizes are between 100 pc and 10 kpc with a peak at around $~$0.9 kpc. Again, in Fig.~\ref{fig:ISMIlustration} we sketch, as an example, the DIG as a volumetric region around the neutral clouds.

\begin{figure*}[ht]
 \includegraphics[width=\textwidth]{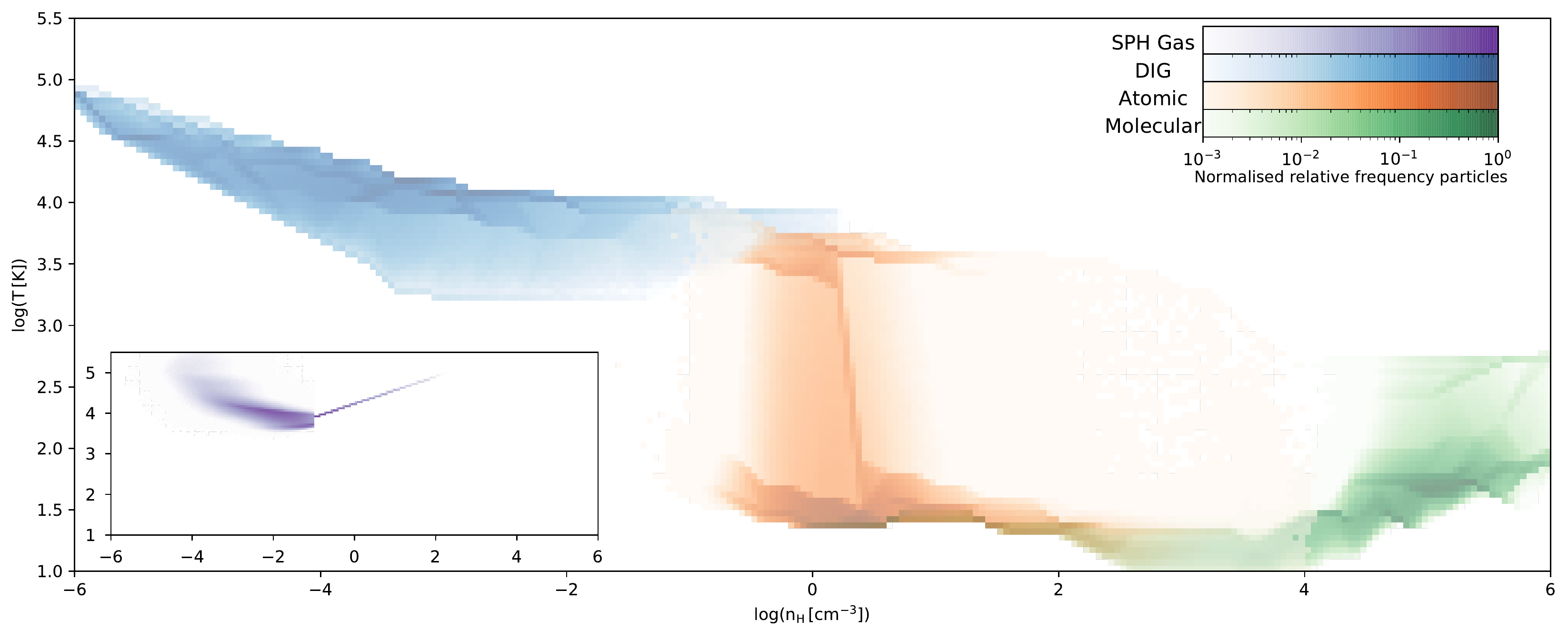}
 \caption{Temperature vs density for the three ISM phases in the \textsc{Recal-L0025N0752} galaxies. Each phase is represented with a normalised relative frequency between 0.1\% and 100\%. We observe that each phase is located in a specific region of the temperature vs density plane. The DIG (blue) is characterised by high temperatures ($\log{T}[\mathrm{K}]\gtrsim3.5$) and low densities ($\log{\mathrm{n(\mathrm{H})}} [\mathrm{cm}^{-3}] \lesssim -1$), atomic gas (orange) by intermediate densities ($-1 \lesssim \log{\mathrm{n(\mathrm{H})}}[\mathrm{cm}^{-3}] \lesssim 3$) with a bridge between high temperatures ($\log{T}[\mathrm{K}] \approx 3.5$) and low temperatures ($\log{T}[\mathrm{K}] \approx 1.4$) and molecular gas (green) by high densities ($\log{\mathrm{n(\mathrm{H})}}[\mathrm{cm}^{-3}]  \gtrsim 3$) and a range of very low temperatures ($\log{T}[\mathrm{K}] \approx 1.1$) to intermediate temperatures ($\log{T}[\mathrm{K}] \approx 2.5$) due to $\mathrm{H_2}$ heating processes \citep{2019ApJ...881..160B}. The temperature-density relation of the original SPH gas particles (inset plot) is shown in purple to demonstrate the need to use a more detailed ISM model when predicting FIR line strengths from these simulations.}
 \label{fig:Densities}
\end{figure*}

In Fig.~\ref{fig:Densities} we show the densities and temperatures obtained with the gas particles for the different ISM phases, as well as the initial SPH gas and density from the simulation. Most of the initial gas density fills the area above 10$^{4}$K and below $\sim 0.1$ cm$^{-3}$, while the remaining initial gas is distributed along the equation-of-state threshold imposed by \textsc{EAGLE}. The initial gas distribution in the temperature-density plane shows how important it is to implement a physically motivated model for the different ISM phases where \textsc{EAGLE} is incapable of reaching. With our model, the DIG density runs from the lowest density (10$^{-6}$) to $\sim 3$ cm$^{-3}$ with a peak around 0.01 cm$^{-3}$, the atomic density runs from  10$^{-1}$ to  10$^{3.5}$ cm$^{3}$ with a peak at 1 cm$^{-3}$, and the molecular gas density runs from 10 cm$^{-3}$ to 10$^{6}$ cm$^{-3}$ with a peak at 10$^{5}$ cm$^{-3}$. For the temperatures, the DIG range is between 10$^{3.2}\,\mathrm{K}$ and 10$^{4.9}\,\mathrm{K}$, with a peak around 10$^{4}\,\mathrm{K}$, in the atomic the temperatures vary from 10$^{1}\,\mathrm{K}$ and 10$^{4}\,\mathrm{K}$ with two peaks around $60\,\mathrm{K}$ and $5000\,\mathrm{K}$, and the molecular gas temperature is constrained to a small region between 10 to $300\,\mathrm{K}$ due to the $\mathrm{H_2}$ heating processes.

Comparing Fig.~\ref{fig:Densities} with recent simulations of individual local-like galaxies (MW and M51) from \citet{2020MNRAS.492.2973T,2020MNRAS.tmp.2910T}, we find similar locations for the atomic gas phases and thermal stable regimes ($T\sim100\,\mathrm{K}$ and $T\sim 10^4\,\mathrm{K}$). Molecular regions are located in the same regime ($T\sim20\,\mathrm{K}$ and $n(\mathrm{H})>10^2\,\mathrm{cm^{-3}}$) as well as diffuse ionised gas ($T\sim10^4\,\mathrm{K}$ and $n(\mathrm{H})<1\,\mathrm{cm^{-3}}$). The only difference is on the step transition we have between the thermal stable regimes. With our \textsc{Cloudy} grids, the temperatures change from $10^4\,\mathrm{K}$ to $100\,\mathrm{K}$ over less than one order of magnitude in density, while \citet{2020MNRAS.492.2973T,2020MNRAS.tmp.2910T} results show a smooth transition that covers two orders of magnitude in density. This transition depends on the assumed metallicity and radiation field \citep{2019ApJ...881..160B}. Therefore, the difference between our results and \citet{2020MNRAS.492.2973T,2020MNRAS.tmp.2910T} may be related to different implementations in terms of numerical methods and chemical evolution, affecting the ISRF and metallicity.

\section{Results} \label{sec:results}

\subsection{Contributions of ISM phases to {[\ion{C}{II}]} emission} \label{subsec:ISMPhase}

Various studies report that the atomic phase dominates the fractional contribution of $L_{\mathrm[\ion{C}{II}]}$ in the MW and the local Universe \citep{2014A&A...570A.121P,2019A&A...626A..23C,2020A&A...639A.110A}. For example, \citet{2019A&A...626A..23C} calculate the physical parameters of the Herschel Dwarf Galaxy Survey (HDGS) sample with a mix of two models, for low and high ionisations. They assume that only the dense \ion{H}{II} model (high ionisation) produces an atomic gas phase. In terms of their atomic gas fraction there is an average increase in their sample from $\sim0.2$ to $\sim0.5$ when $\log(\mathrm{SFR})\,[\mathrm{M}_\sun \mathrm{yr}^{-1}]$ changes from $\sim -3$ to $\sim0$. \citet{2014A&A...570A.121P} calculated the ISM contributions in the MW galactic plane where molecular, atomic, and ionised gas, contribute to 25\%, 55\%, and 20\% of the {[\ion{C}{II}]} luminosity, respectively. However, the contribution of the DIG phase is difficult to estimate accurately. For example, the DIG phase can be more vertically extended in disk galaxies such as the MW, which can cause a difference in the estimated DIG contribution from 4\% to 20\% as noted by \citet{ 2013A&A...554A.103P,2014A&A...570A.121P}. Finally, \citet{2020A&A...639A.110A} studied the Orion-Eridanus super-bubble and found that the surfaces of molecular clouds, mostly atomic gas, are the main contributors (80\%) to the {[\ion{C}{II}]} emission.

\begin{figure}[ht]
\includegraphics[width=0.98\columnwidth,height=0.8\textheight,keepaspectratio]{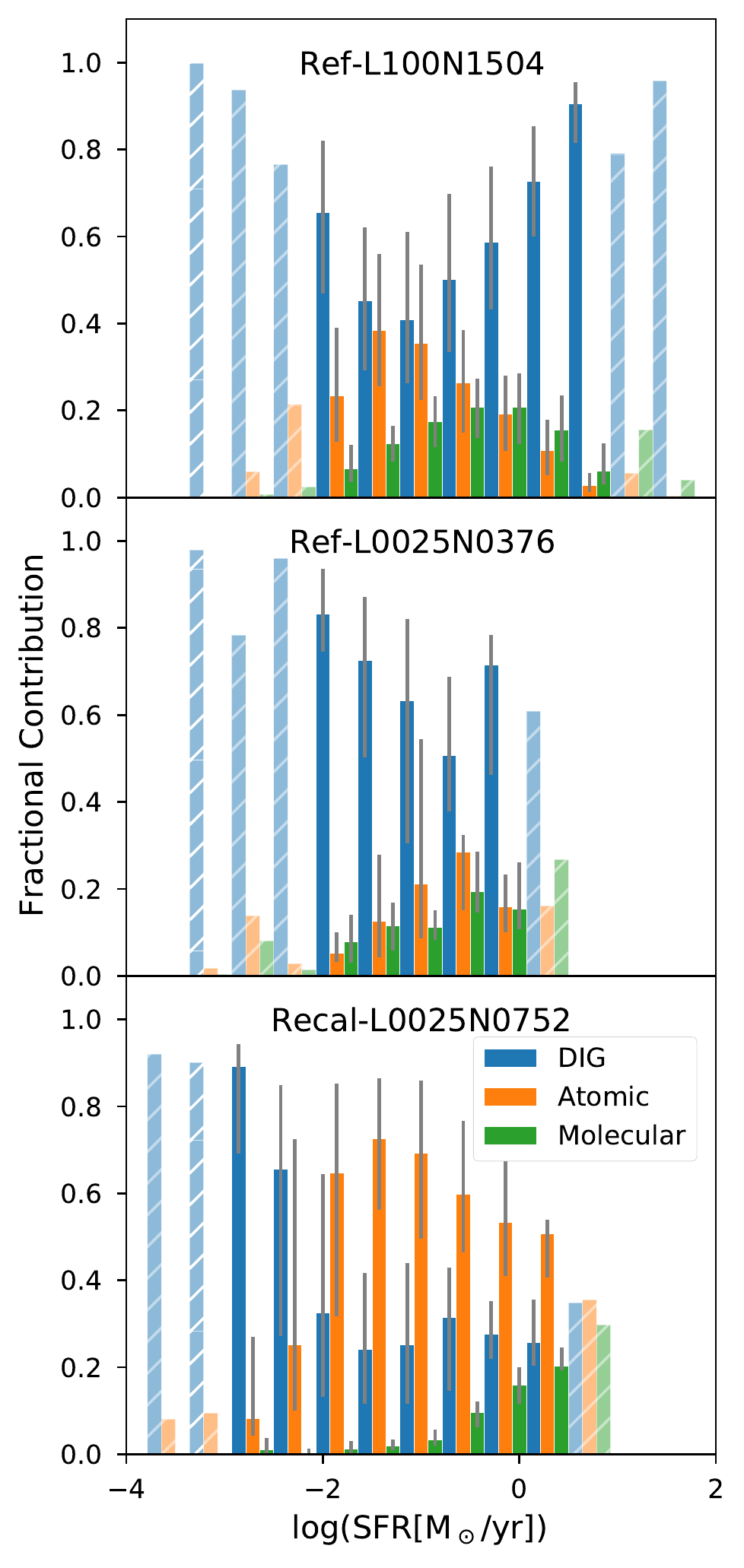}
 \caption{Fractional contributions of the ISM phases to the total $L_{\mathrm[\ion{C}{II}]}$ for \textsc{Ref-L0100N1504} (\textit{top panel}), \textsc{Ref-L0025N0376} (\textit{middle panel}), and \textsc{Recal-L0025N0752} (\textit{bottom panel}) simulations. Shaded, white-diagonal-striped bars indicate bins with fewer than ten galaxies. The grey lines show the range between the 25th and 75th percentiles. Colours are the same as in Fig.~\ref{fig:Densities}.}
 \label{fig:ISMPha}
\end{figure}

We separate the contributions from different ISM phases to $L_{\mathrm[\ion{C}{II}]}$ as a function of SFR. We present the median fractional contribution (the luminosity contribution of a given phase to the total luminosity of the galaxy) of these ISM phases in Fig.~\ref{fig:ISMPha} for the three simulations we have used. These results show the fractional contribution of each of the ISM phases presented in Fig.~\ref{fig:ISMPha} varies with the simulation used and the total SFR of the galaxy. 

We see a general agreement in terms of qualitative trends among the three simulations in Fig.~\ref{fig:ISMPha}. In all simulations, the fractional contributions from the neutral phases increase with increasing SFR, especially for the molecular phase. On the other hand, the DIG fractional contribution decreases with increasing SFR and then flattens, or increases again for \textsc{Ref-L0100N1504}. However, the three simulations also exhibit some differences in quantitative terms. The relatively small differences between the two intermediate simulations, \textsc{Ref-L0025N0376} and \textsc{Ref-L0100N1504}, are due to sampling noise as the two simulations only differ in size. In particular, the larger box simulation provides a better sampling of the brighter and rarer galaxy population with higher SFRs. The differences between the intermediate and high-resolution simulations are much larger and are due to the GSMF re-calibration, which changes the feedback parameters in the simulation \citep{2015MNRAS.446..521S}. This re-calibration leads to different intrinsic characteristics for the galaxies between the simulations: The stellar feedback produces more outflows of metal-enriched material in \textsc{Recal-L0025N0752} (compared to \textsc{Ref-L0100N1504} and \textsc{Ref-L0025N0376}), reducing the total metallicity in these galaxies \citep{2015MNRAS.446..521S, 2017MNRAS.472.3354D}. In addition, the effect of AGN feedback is important for changing the metallicities of high-mass galaxies, as pointed out by \citet{2017MNRAS.472.3354D}. Changes in metallicity are responsible for the differences in the $L_{\mathrm[\ion{C}{II}]}$ estimates between these simulations.

We can compare the fractional contributions in \textsc{Recal-L0025N0752} with observational studies where most of the {[\ion{C}{II}]} comes from the atomic phase instead of the DIG. In \textsc{Recal-L0025N0752} the atomic phase becomes increasingly more important with increasing SFR like in \citet{2019A&A...626A..23C}, with smaller contributions from the dense molecular gas and the DIG.  However, in this work we do not model the \ion{H}{II} regions, so our results are not entirely comparable with \citet{2019A&A...626A..23C}. Assuming a $\log(\mathrm{SFR})\,[\mathrm{M}_\sun \mathrm{yr}^{-1}]\sim0$ for the MW \citep{2010ApJ...710L..11R}, the ISM contributions from \textsc{Recal-L0025N0752} at this SFR are $14_{-2}^{+4}$\%, $55_{-13}^{+13}$\% and $25_{-3}^{+12}$\% for the molecular, atomic and ionised gas, similar to the fractions presented by \citet{2014A&A...570A.121P}. Similar results are found in nearby dwarf galaxies \citep{2017A&A...599A...9F,2019A&A...626A..23C} and spiral galaxies \citep{ 2017ApJ...842....4A}. Thus, results from \textsc{Recal-L0025N0752} agree well with observations in the local Universe.

\subsection{The [\ion{C}{II}]--SFR relation} \label{subsec:CII-SFR}

\begin{figure*}
 \includegraphics[width=\textwidth]{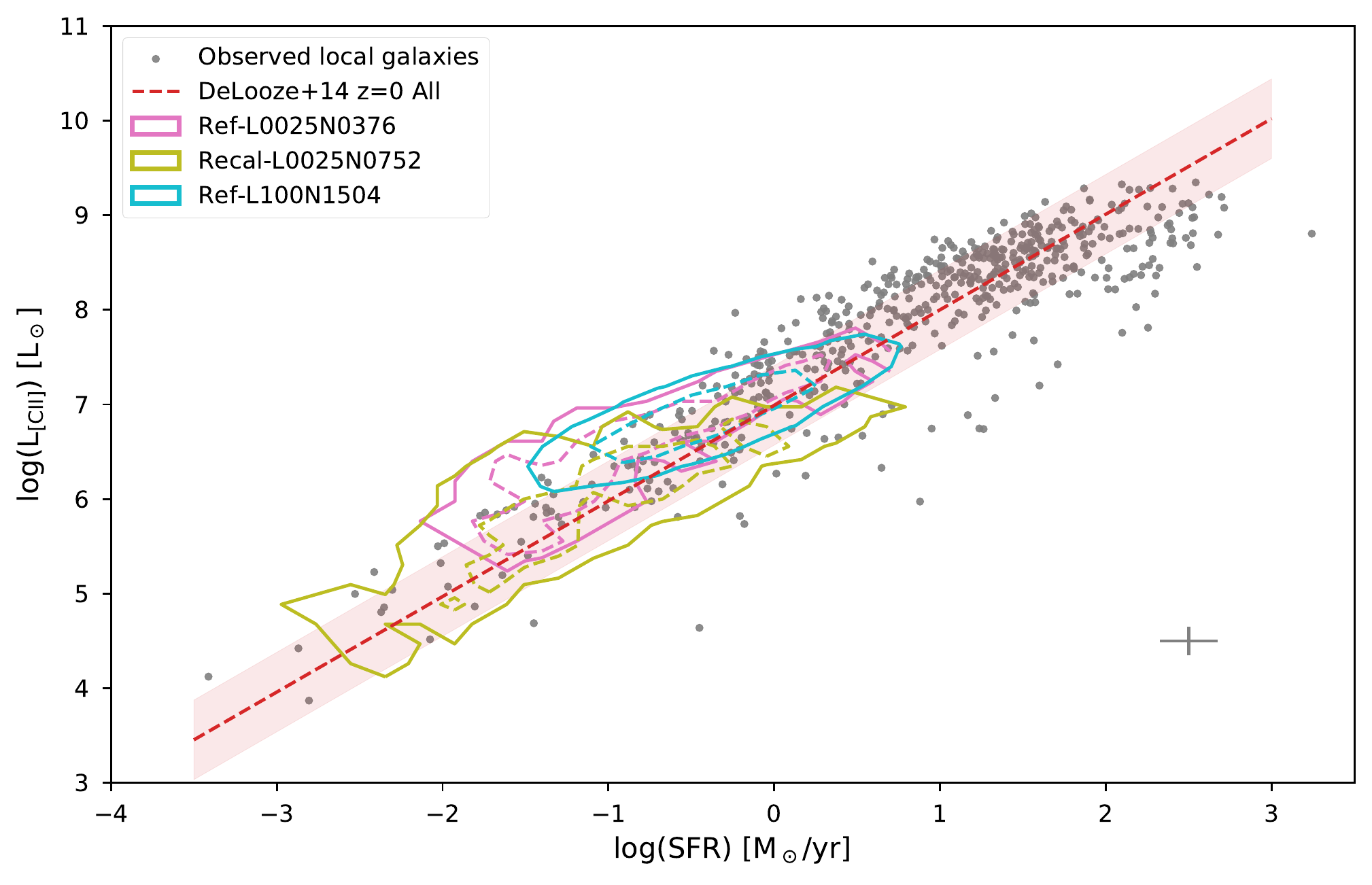}
 \caption{$L_{\mathrm[\ion{C}{II}]}$--SFR relation for the three simulations used in this work (\textsc{Ref-L0025N0376}, \textsc{Recal-L0025N0752}, and \textsc{Ref-L0100N1504}) presented as contour maps (pink, olive and cyan, respectively) and an observational sample of local galaxies (grey dots, briefly described in Appendix~\ref{App:Observations}). The contours shows where 90\% (solid) and 50\% (dashed) of the galaxies of the respective simulations fall in the relation. The sample of local galaxies is a compilation of different surveys containing main sequence galaxies \citep[ISO Compendium]{2008ApJS..178..280B}, starburst galaxies \citep[GOALS]{2013ApJ...774...68D,2017ApJ...846...32D}, dwarf galaxies \citep[HDGS]{2015A&A...578A..53C,2019A&A...626A..23C}, star-forming, AGN and LIRG galaxies \citep[SHINING]{ 2018ApJ...861...94H,2018ApJ...861...95H}, dusty main sequence galaxies \citep[VALES]{2017A&A...602A..49H} and intermediate-stellar-mass galaxies \citep[xCOLD GASS]{2017MNRAS.470.4750A}. We present the mean error from the observational samples in the bottom-right corner of the plot. The dashed red line is the power law derived by \citet{2014A&A...568A..62D} for the relation at $z=0$, with the shaded red region representing the 1$\sigma$ scatter.}
 \label{fig:z0}
\end{figure*}

In this section, we compare the well-known relation between $L_{\mathrm[\ion{C}{II}]}$ and SFR in galaxies \citep{2014A&A...568A..62D} with that obtained with our model. As we mentioned before (Sect.~\ref{sec:intro}), {[\ion{C}{II}]} luminosity can be used as a SFR indicator \citep[e.g.][]{1991ApJ...373..423S,2010ApJ...724..957S,2001ApJ...561..766M}. However, at higher IR luminosities the {[\ion{C}{II}]} luminosities are lower than expected, this effect is known as the `{[\ion{C}{II}]} deficit'. Various reasons has been proposed \citep{2016MNRAS.463.2085M,2017MNRAS.467...50N,2017ApJ...834....5S,2017ApJ...846...32D,2019MNRAS.489....1F} pointing mainly to local conditions of the interstellar gas as the drivers of this deficit, such as the intensity of the radiation field, metallicity or gas density. However, the impact from other physical parameters such as dust temperatures \citep{2018ApJ...869L..30P} or the AGN \citep{2018ApJ...861...95H} are still not clear. In Fig.~\ref{fig:z0} we see the {[\ion{C}{II}]} deficit when we compare the observational sample with the power law derived by \citet{2014A&A...568A..62D}: at higher SFR $\gtrsim 10\,\mathrm{M_{\sun}}\,\mathrm{yr^{-1}}$ the slope with $L_{\mathrm[\ion{C}{II}]}$ is less steep than at low SFR $\lesssim 10\,\mathrm{M_{\sun}}\,\mathrm{yr^{-1}}$ (FIR luminosities $\lesssim 10^{11} \mathrm{L_{\sun}}$). The $L_{\mathrm[\ion{C}{II}]}$--SFR relation is complex and a simple power law (like the one implemented by \citet{2014A&A...568A..62D}) cannot describe it \citep{2019MNRAS.489....1F}. This trend is also observed in other FIR lines as presented by \citet{2017ApJ...846...32D}. 

The $L_{\mathrm[\ion{C}{II}]}$--SFR relation we inferred from the ISM model is presented in Fig.~\ref{fig:z0} and compared with selected samples of observed galaxies in the local Universe. We briefly describe the observational samples that we compare to in this paper in Appendix~\ref{App:Observations}. The multi-phase ISM model we have implemented in this work reproduces the observed galaxy distribution in the $L_{\mathrm[\ion{C}{II}]}$--SFR relation from \citet[][entire sample]{2014A&A...568A..62D} and \citet{2014A&A...570A.121P}. In the range $-3.5<\log(\mathrm{SFR})\,[\mathrm{M}_\sun \mathrm{yr}^{-1}]<3$, 77\% of \textsc{Recal-L0025N0752} galaxies are inside the 1$\sigma$ dispersion of the \citet{2014A&A...568A..62D} relation. For the intermediate-resolution simulations (\textsc{Ref-L0025N0376} and \textsc{Ref-L0100N1504}), only 47--60\% of galaxies are inside the 1$\sigma$ dispersion. If we compare with the 2$\sigma$ dispersion around the \citet{2014A&A...568A..62D} relation, we find 83, 95 and 96\% of the galaxies inside the dispersion for \textsc{Ref-L0025N0376}, \textsc{Ref-L0100N1504} and \textsc{Recal-L0025N0752}, respectively. The dispersion (1$\sigma$) of the simulations (around 0.4 dex) is comparable to the \citet{2014A&A...568A..62D} relation (0.42 dex). The typical statistical uncertainty in the observed SFR ($L_{\mathrm[\ion{C}{II}]}$) is around 10\% (5\%) but can go up to 40\% (35\%) in some cases \citep[e.g.][]{2015A&A...578A..53C,2017MNRAS.470.3775V,2017ApJ...846...32D}. 

The simulations underestimate the abundance of star-forming galaxies ($1-10 \,\mathrm{M_{\sun}}\,\mathrm{yr^{-1}}$) in the local Universe due to the AGN feedback implementations used by \textsc{EAGLE} \citep{2017MNRAS.472..919K}.
In addition, the relatively small volume and physical prescriptions of the simulations (e.g. the IMF, lack of starburst galaxies and sub-grid physics) limit the comparison with luminous IR galaxies (e.g. (U)LIRGs) with SFR above $10\,\mathrm{M_{\sun}}\,\mathrm{yr^{-1}}$ \citep{2019A&A...624A..98W}. Thus, the results of this work are mainly for galaxies with SFRs below $10\,\mathrm{M_{\sun}}\,\mathrm{yr^{-1}}$. However, caution is need when comparing theoretical results with observational measurements as there could be important systematic differences. For example, the SFRs in EAGLE are computed from the KS law \citep{2008MNRAS.383.1210S}, while observational measurements of SFRs are normally based on IR luminosity.

Overall, Figs.~\ref{fig:ISMPha} and ~\ref{fig:z0} imply that in our simulated galaxies, the dominant phase of $L_{\mathrm[\ion{C}{II}]}$ depends on the star-formation activity of the galaxy and the impact on each of the ISM phases can explain the shape of the $L_{\mathrm[\ion{C}{II}]}$--SFR relation (see discussions in Sects.~\ref{subsec:DisDIG} and ~\ref{subsec:DisDeficit}). Although the simulations do not go to SFR $\sim 100\,\mathrm{M_{\sun}}\,\mathrm{yr^{-1}}$, we can study the variations and trends of the physical parameters, which can provide some insights into the {[\ion{C}{II}]} deficit (Sect.~\ref{subsec:DisDeficit}).

\subsection{Pressure and metallicity dependence} \label{subsec:PreMet}

To understand the origin of the $L_{\mathrm[\ion{C}{II}]}$--SFR relation, it is important to study how this relation varies as a function of other physical parameters. In this section, we study its dependence on pressure and metallicity. 

\begin{figure}[ht]
  \includegraphics[width=\columnwidth]{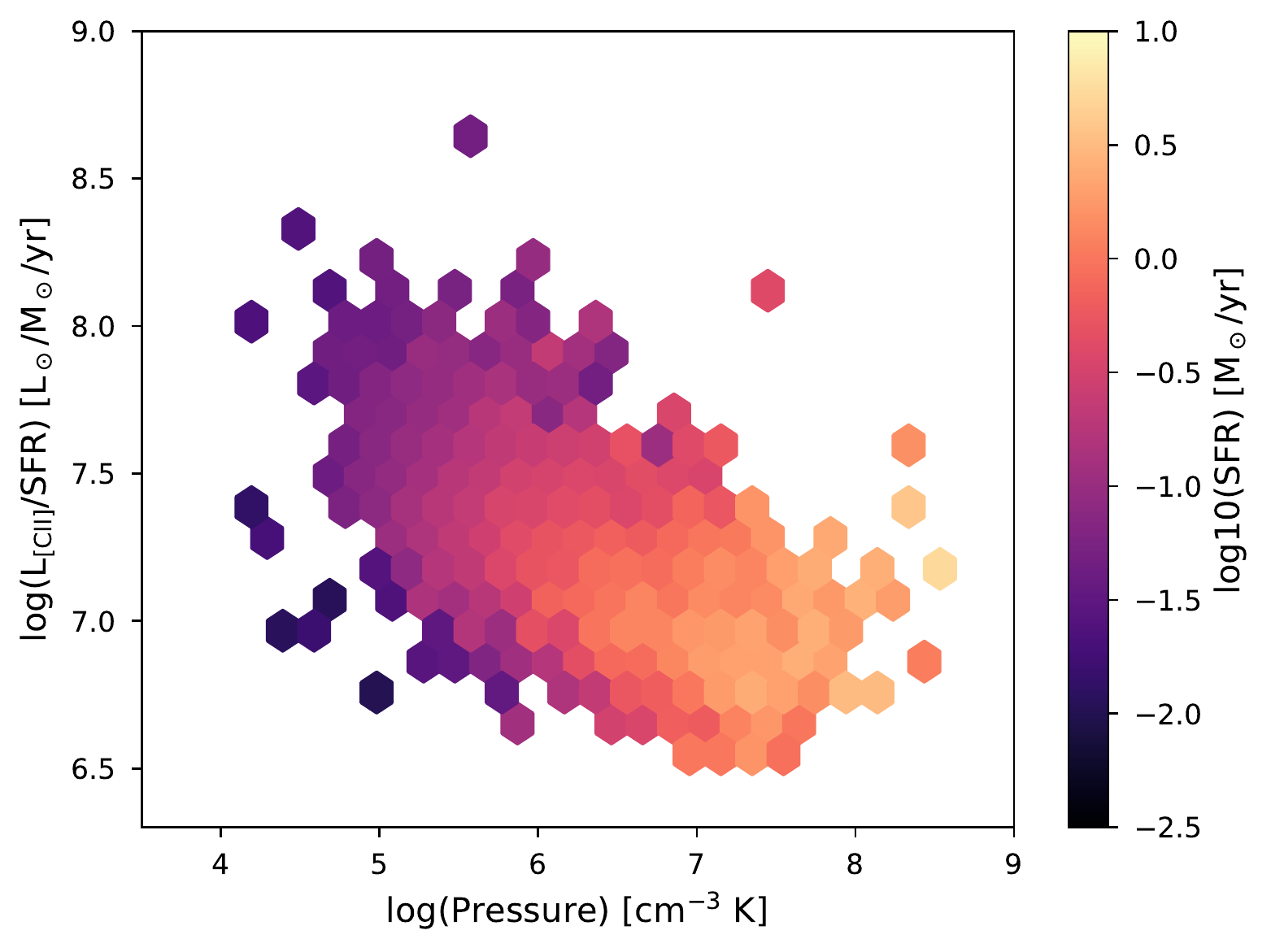}
 \caption{The $L_{\mathrm[\ion{C}{II}]}$/SFR ratio as a function of pressure, colour-coded by SFR. We present the mean value of SFR for each bin for the galaxies in the three simulations assuming they represent the same population. Pressure is anti-correlated with the $L_{\mathrm[\ion{C}{II}]}$/SFR ratio, in agreement with the result presented by \citet{2019MNRAS.482.4906P}. We have confirmed that the trends do not depend on the specific simulation used.}
 \label{fig:PresMet}
\end{figure}

We plot the $L_{\mathrm[\ion{C}{II}]}$/SFR ratio as a function of pressure in Fig.~\ref{fig:PresMet}. We combine the three simulations. We bin the pressure and $L_{\mathrm[\ion{C}{II}]}$/SFR values with bin sizes determined using Knuth’s rule \citep{2006physics...5197K}, which selects the simplest model that best describes the data by maximising the posterior probability for the number of bins. We colour-code the hexagonal bins by SFR in the range $-2.5<\log(\mathrm{SFR})\,[\mathrm{M_{\sun}}\,\mathrm{yr^{-1}}]<1$. We obtain similar results as \citet{2019MNRAS.482.4906P}: At fixed SFR, the ratio of $L_{\mathrm[\ion{C}{II}]}$/SFR decreases with increasing pressure. This trend is expected 
as the  mass-size relation of the neutral clouds in the model (Eq.~\ref{eq:GMCMassSize}) depends on the pressure. At high pressure, the cloud is smaller and the emission region from the {[\ion{C}{II}]} neutral phase decreases. We discuss the neutral cloud size dependency in Sect.~\ref{subsec:DisDeficit}.

We use observational metallicity ($12 + \log(\mathrm{O/H})$) values from the HDGS \citep{2019A&A...626A..23C}, xCOLD GASS \citep{2017MNRAS.470.4750A} and VALES samples \citep[taken from][]{2018MNRAS.481.1976Z,2017A&A...602A..49H}. We convert $L_{\mathrm{IR}}$ to SFR as described by \citet{2012ARA&A..50..531K}. As presented by \citet{2008ApJ...681.1183K}, there is a large difference in the type of $12 + \log(\mathrm{O/H})$ calibration used, in some cases up to 0.7 dex in $\log(\mathrm{O/H})$. For example, VALES and xCOLD GASS samples use the calibration from \citet{2004MNRAS.348L..59P}, while the HDGS sample uses a compilation of metallicities \citep{2013A&A...557A..95R}. In this work, we do not calibrate the observations to a single metallicity calibration. In EAGLE, gas metallicity is defined as the fraction of the gas mass in elements heavier than helium. We use a solar metallicity of $12 + \log(\mathrm{O/H}) = 8.69$ or $Z=0.0134$ \citep{2009ARA&A..47..481A} to compare EAGLE and observational values.

\begin{figure*}[ht]
 \includegraphics[width=\textwidth]{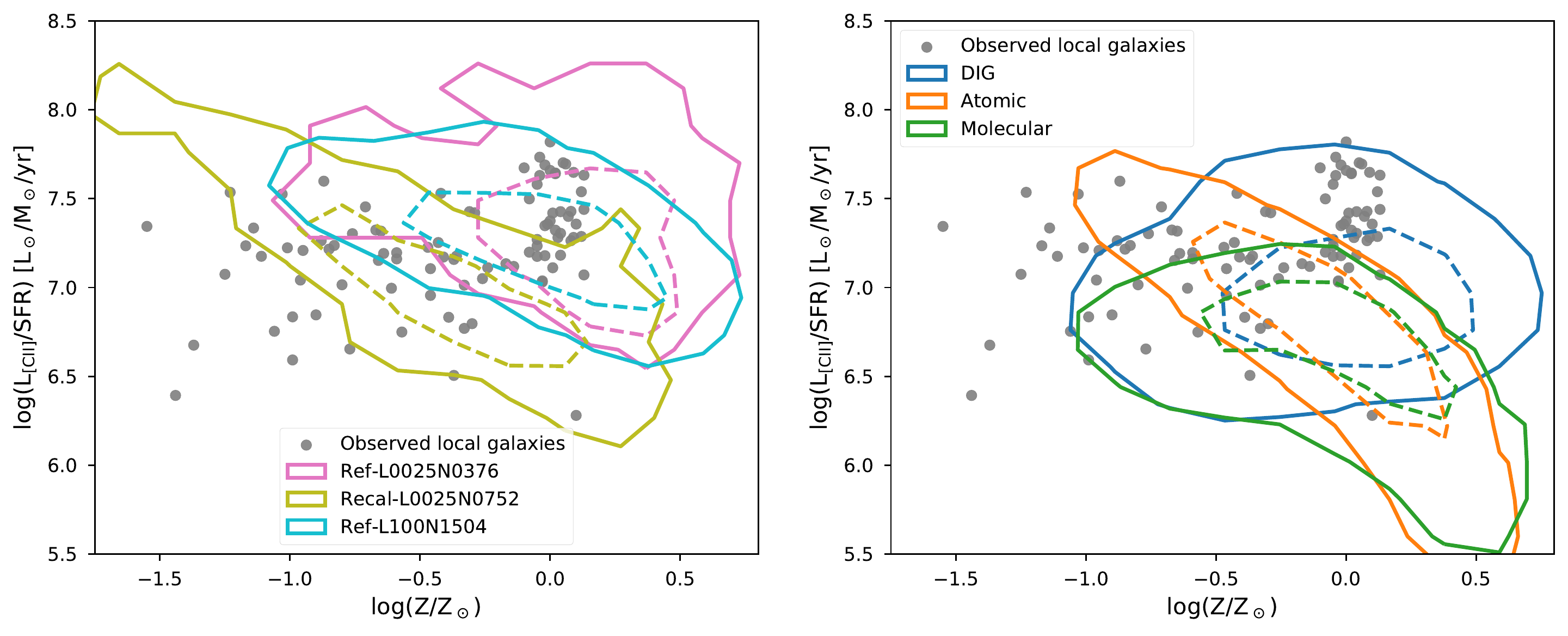}
 \caption{The $L_{\mathrm[\ion{C}{II}]}$/SFR ratio as a function of metallicity. \textit{Left panel}: Simulations compared with observed galaxies. Contours show where 90\% (solid) and 50\% (dashed) of the galaxies of the respective simulation lie. \textit{Right panel}: The different ISM phases (DIG, atomic and molecular) in \textsc{Ref-L0100N1504}. For the atomic phase, the $L_{\mathrm[\ion{C}{II}]}$/SFR increases with decreasing metallicity. For the molecular phase, it also decreases with increasing metallicity but plateaus at $\sim1\,Z_{\sun}$. For the DIG phase, it remains more or less constant.}
 \label{fig:ZEvol}
\end{figure*}

We show the variation of the $L_{\mathrm[\ion{C}{II}]}$/SFR ratio with metallicity for the three simulations in Fig.~\ref{fig:ZEvol}. There is no clear trend in the observational data, as we only have a few data points. The $L_{\mathrm[\ion{C}{II}]}$/SFR ratio in the simulations show a strong dependence on the mean gas metallicity values. Galaxies in \textsc{Ref-L0025N0376} and \textsc{Ref-L0100N1504} are located at higher values from the $L_{\mathrm[\ion{C}{II}]}$/SFR ratio compared to \textsc{Recal-L0025N0752} (as in Fig.~\ref{fig:z0}), due to a higher contribution of the DIG phase (see discussion in Sect.~\ref{subsec:DisRightSim}). There is a decrease in the $L_{\mathrm[\ion{C}{II}]}$/SFR ratio from low metallicities up to $3\,Z_{\sun}$ in galaxies in \textsc{Recal-L0025N0752}. The total number of observed galaxies and simulated galaxies in the low-metallicity regime (below $\sim0.2\,Z_{\sun}$) is not enough to establish a clear trend between metallicity and $L_{\mathrm[\ion{C}{II}]}$/SFR as presented by \citet{2015ApJ...813...36V} and \citet{2020MNRAS.492.2818L}, where there is an increase in the $L_{\mathrm[\ion{C}{II}]}$/SFR ratio with metallicity.

For metallicities higher than solar, we see a clear reduction in the $L_{\mathrm[\ion{C}{II}]}$/SFR ratio in \textsc{Ref-L0100N1504}; therefore, we plot in the right panel of Fig.~\ref{fig:ZEvol} the different ISM phases in \textsc{Ref-L0100N1504} as a function of metallicity. For the atomic phase, the $L_{\mathrm[\ion{C}{II}]}$/SFR ratio decreases with increasing metallicity.  For the molecular phase, the ratio is more or less  flat below solar metallicity, and then decreases with increasing metallicity above solar metallicity. On the other hand, the DIG is flat in the $L_{\mathrm[\ion{C}{II}]}$/SFR ratio at all metallicities. The decrease in the $L_{\mathrm[\ion{C}{II}]}$/SFR ratio at high metallicities can make a significant impact on the observed total $L_{\mathrm[\ion{C}{II}]}$ when the atomic and molecular phases dominate $L_{\mathrm[\ion{C}{II}]}$. This probably results from higher radiation fields in metal-rich galaxies, which reduce the sizes of the neutral clouds, as proposed by \citet{2017MNRAS.467...50N}. We find that galaxies with high metallicity have average neutral cloud sizes $\sim30$\% smaller than metal-poor galaxies. We discuss this further in Sect~\ref{sec:disc}. 

\subsection{Spatial distributions} \label{subsec:SpatialDist}

\begin{figure}[ht]
 \includegraphics[width=\columnwidth]{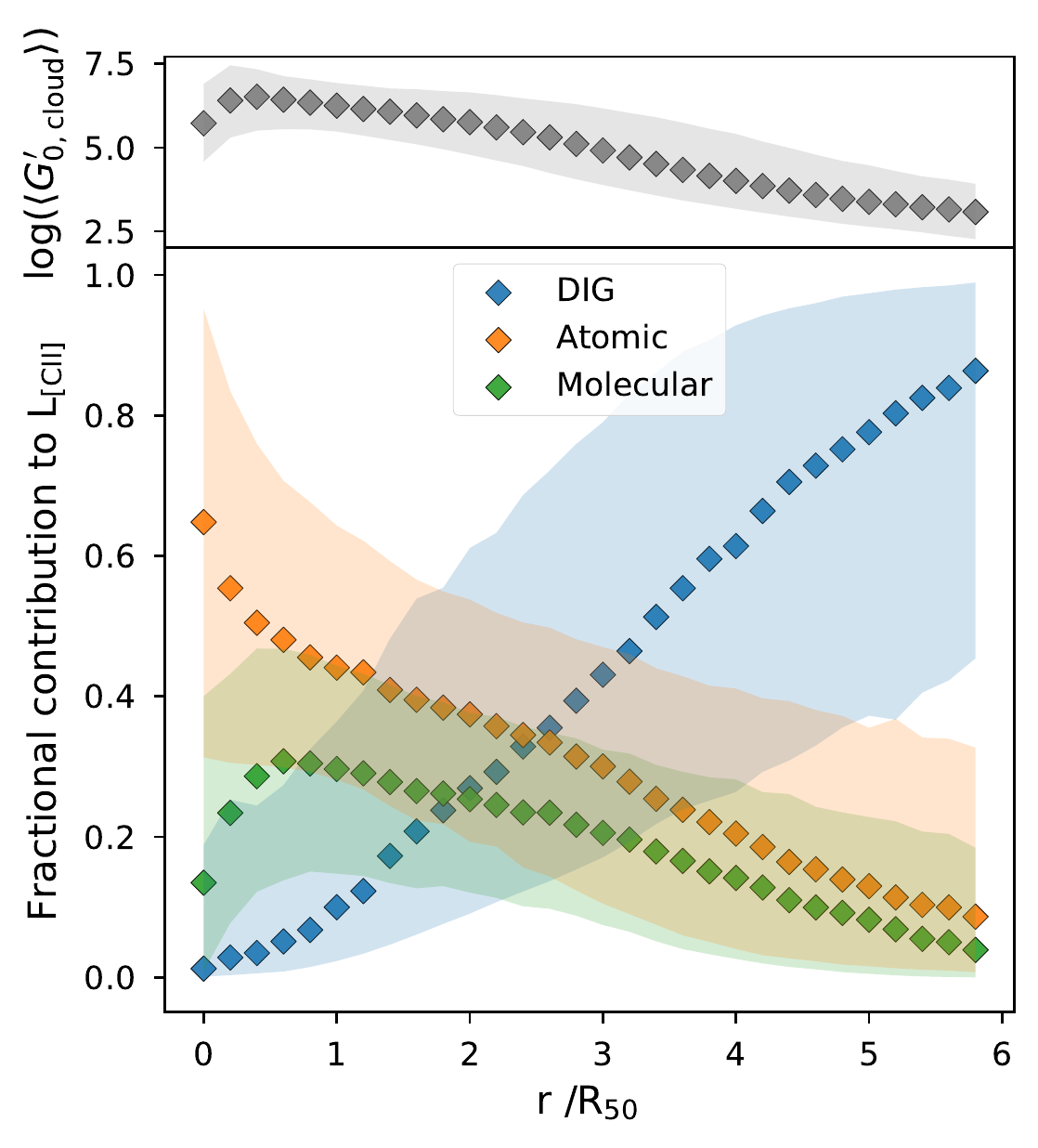}
 \caption{Median radial distribution (diamonds) of the radiation field (grey) in the neutral clouds (\textit{upper panel}) and the fractional ISM phase contributions (\textit{lower panel}) for galaxies in \textsc{Ref-L0100N1504}. There is a peak of neutral atomic gas contribution in the centre, the DIG contribution dominates beyond three times the half-mass radius ($R_{50}$) of the galaxies, and the dense molecular gas contribution peaks around $0.5R_{50}$ and then decreases slowly. The peak of the radiation field is similar to the one of the dense molecular gas. Shaded colour shows the coverage between the 25th and 75th percentiles.}
 \label{fig:Radial}
\end{figure}

Finally, we check the spatial distribution of the ISM fractional contributions inside the galaxies. We calculate the distance between each phase to the centre of the potential in each galaxy. We scale the distance using the half-mass radius ($R_{50}$) for all galaxies in \textsc{Ref-L0100N1504} from the EAGLE Database. We present the fractional contribution of the ISM phases against the distance from the centre in units of $R_{50}$ in Fig.~\ref{fig:Radial}. These radial distributions of the fractional contributions to $L_{\mathrm[\ion{C}{II}]}$ show that atomic gas dominates the luminosity contribution at the centres of the galaxies. The molecular phase never dominates, but the contribution peaks at r/$R_{50} \approx 0.5$, while in the outskirts the DIG dominates.

The atomic and molecular phases are confined to the galaxy centres compared with the DIG. Atomic gas always dominates below $R_{50}$, while the DIG median is nearly always lower than the molecular contribution within $\sim R_{50}$. The DIG could be significant when $L_{\mathrm[\ion{C}{II}]}$ is measured in observations of unresolved galaxies. As we show in Fig.~\ref{fig:ISMPha}, the contribution of each phase also depends on the SFR of the galaxy. At high SFR, the molecular fraction increases while the atomic contribution decreases. In the same way, the DIG dominates at low SFR, which will lead to a different radial distribution in less actively star-forming galaxies. However, we expect a similar trend where the fractional contributions of atomic and molecular gas peak at the centres of the galaxies and the DIG dominates in the outskirts.

\section{Discussion} \label{sec:disc}

\subsection{The role of the DIG in the {[\ion{C}{II}]} emission of galaxies} \label{subsec:DisDIG}

One of the main conclusions we can draw from the ISM model implemented in this work is the role that the DIG plays in the production of $L_{\mathrm[\ion{C}{II}]}$. We show that the fractional contribution of each major ISM phase depends on the total SFR of the galaxy and the simulation used in Fig.~\ref{fig:ISMPha}. Intermediate-resolution simulations seem to overestimate the DIG contribution, while \textsc{Recal-L0025N0752} is consistent with previous studies in the local Universe \citep[e.g.][]{ 2014A&A...570A.121P,2017MNRAS.467...50N,2017ApJ...845...96C,2019A&A...626A..23C,2020MNRAS.492.2818L}, where most of the {[\ion{C}{II}]} emission arises from neutral phases (atomic and molecular). Figure~\ref{fig:ISMPha} shows that when \textsc{Recal-L0025N0752} is used, the neutral phases dominate the $L_{\mathrm[\ion{C}{II}]}$ fractional contribution, from around 60\% at $\mathrm{SFR}\approx10^{-1}\,\mathrm{M_{\sun}}\,\mathrm{yr^{-1}}$ to 80\% at $\mathrm{SFR}\approx10\,\mathrm{M_{\sun}}\,\mathrm{yr^{-1}}$. Similar results are obtained when we look only at the inner parts of galaxies, inside $R_{50}$ (Fig.~\ref{fig:Radial}). In galaxy centres, the total atomic contribution is $\sim 70\%$ and the molecular contribution is $\sim 20\%$, while at $R_{50}$ the atomic contribution is $\sim 45\%$ and the molecular contribution is $\sim 30\%$.

The DIG is the dominant phase at metallicities above solar ($\sim 70$\% of $L_{\mathrm[\ion{C}{II}]}$), at $R_{50}\gtrsim3$, and in the low-SFR regime ($<10^{-2}\,\mathrm{M_{\sun}}\,\mathrm{yr^{-1}}$) in all three simulations. The $L_{\mathrm{[\ion{C}{II}],DIG}}$/SFR ratio is more or less constant as a function of metallicity for the DIG phase (Fig.~\ref{fig:ZEvol}), in contrast to the atomic and molecular contributions, where the ratio is reduced drastically above solar metallicity. Below 1/3 $Z_\sun$ the atomic gas is the dominant component in all three simulations (up to 90\% at $Z/Z_\sun=0.03$), while the contribution of the dense molecular gas peaks near to solar metallicity. Thus the DIG is the dominant contributor to $L_{\mathrm[\ion{C}{II}]}$ in high-metallicity galaxies. These results support the trend presented by other works  \citep{2017ApJ...845...96C,2019A&A...626A..23C}, where the fractional contribution of the DIG increases at high metallicities. 

The radial position where the fractional contribution of the DIG is high corresponds to the outskirts of galaxies (Fig.~\ref{fig:Radial}). This agrees visually with the expected DIG location from \citet[see their Appendix A]{2019MNRAS.484.5009E}, who estimated the location of the DIG regions, following a threshold value for the H$\alpha$ emission (diffuse H$\alpha$ emission), mainly in the outskirts of local star-forming galaxies. Although they do not compare H$\alpha$ with {[\ion{C}{II}]} emission, it is interesting to note the resemblance with another SFR tracer. This suggests that it could be possible to measure the DIG contribution in the outskirts of galaxies by measuring the {[\ion{C}{II}]} flux, just like H$\alpha$ emission. \citet{2017ApJ...845...96C} and \citet{2019ApJ...886...60S} found fractional contributions of DIG below 40\% in resolved regions of galaxies, which coincide with our results inside the half-mass radii of galaxies (2--6 kpc). However, the results of \citet{2017ApJ...845...96C} and \citet{ 2019ApJ...886...60S} come mostly from regions within $0.25R_{25}$\footnote{1/8 of the $D_{25}$ standard diameter, the $B$-band isophotal radius at 25 mag arcsec$^{-2}$ \citep{1991rc3..book.....D, 1991A&A...243..319P,1998A&AS..128..299P}.}, limiting the interpretation of these studies on the spatial distribution of the DIG phase.

In an analytical study, \citet{2019MNRAS.482.4906P} assume a constant density for diffuse gas. When \citet{2019MNRAS.482.4906P} reduce the density of the diffuse gas, the total {[\ion{C}{II}]} luminosity is more affected (up to 0.5 dex with respect to their fiducial model) at $\log(\mathrm{SFR})\,[\mathrm{M_{\sun}}\,\mathrm{yr^{-1}}]<0$ than at $\log(\mathrm{SFR})\,[\mathrm{M_{\sun}}\,\mathrm{yr^{-1}}]>0$ (their Fig. 15). This result is consistent with the trends found in our work. When the DIG dominates at low SFRs, even small changes in the DIG composition can affect the total {[\ion{C}{II}]} luminosity. However, the assumption of a constant density for diffuse gas and other differences in the ISM models, limits a direct comparison of the DIG dependency on $L_{\mathrm[\ion{C}{II}]}$. 

Our results show that, in metal-rich galaxies and where the SFR is low, the DIG could play a dominant role in producing the $L_{\mathrm[\ion{C}{II}]}$. Ignoring these contributions at different SFRs can introduce a bias in the line emission estimations \citep{2017ApJ...845...96C}. Detailed resolved observations of local galaxies and their outskirts will be essential to improve the calibration of the DIG fraction. At the same time, observations of galaxies with low SFR will also be important. 

\subsection{Variations in the $L_{\mathrm[\ion{C}{II}]}$--SFR relation} \label{subsec:DisDeficit}

As mentioned in Sects.~\ref{sec:intro} and \ref{subsec:CII-SFR}, one problem of using $L_{\mathrm{[\ion{C}{II}]}}$ as a SFR indicator is that the $L_{\mathrm[\ion{C}{II}]}$--SFR relation is complex. A single power law cannot describe the relation accurately and variations are present due to changes in contribution from different ISM phases (Sects.~\ref{subsec:ISMPhase} and~\ref{subsec:DisDIG}) and the {[\ion{C}{II}]} deficit, where the slope of $L_{\mathrm[\ion{C}{II}]}$ with respect to FIR luminosity is less steep at high infrared luminosities. 

\begin{figure}
 \includegraphics[width=\columnwidth]{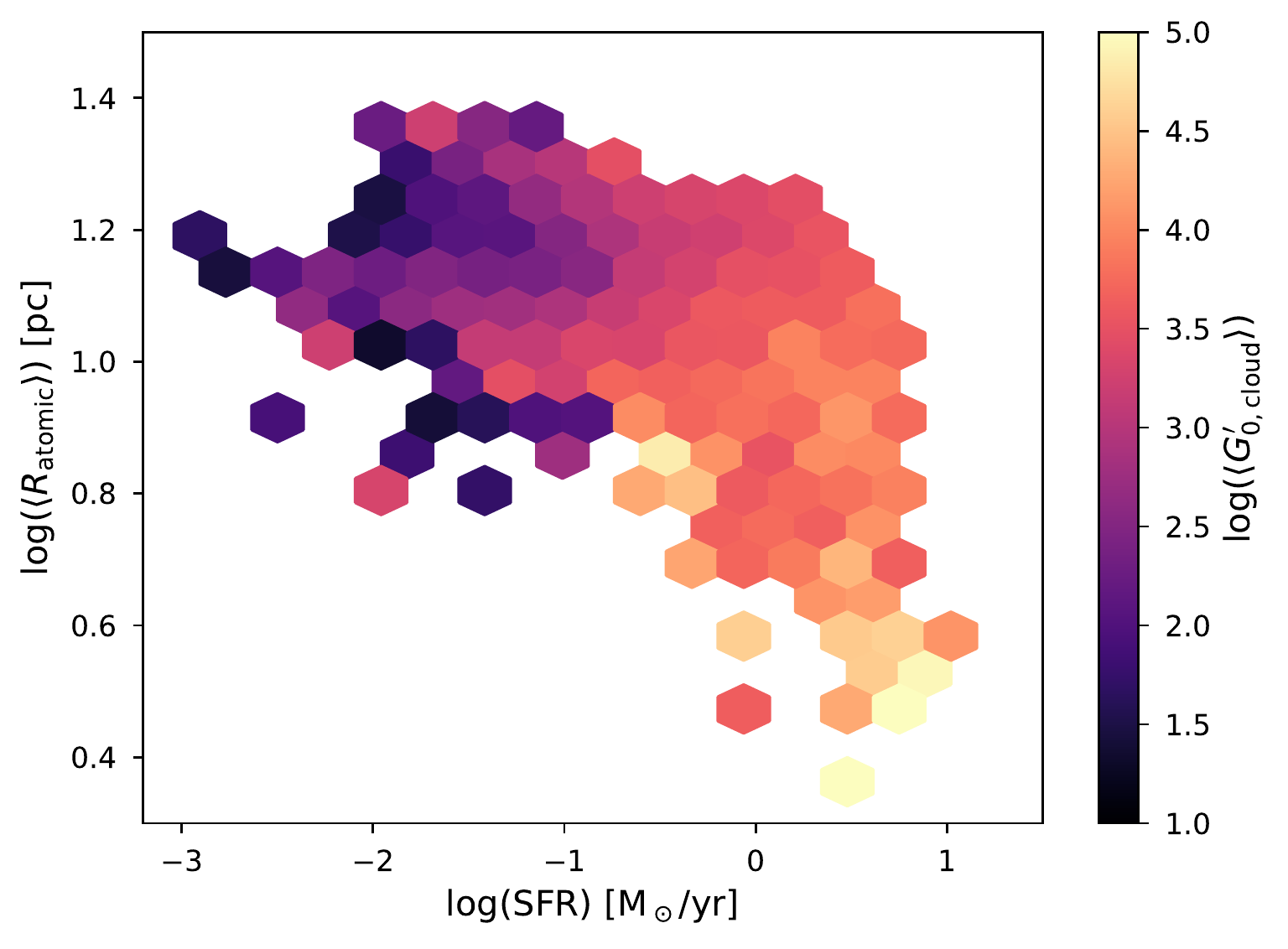}
 \caption{Average radius of the atomic region ($R_{\mathrm{atomic}}$) in all simulations (assuming they represent the same population) with respect to SFR, colour-coded by $G_{0}^{\prime}$ (Habing units). We binned the $R_{\mathrm{atomic}}$ values to bin sizes of $\sim0.05$ dex and the SFR to sizes of $\sim0.2$ dex. This plot supports the idea proposed by \citet{2017MNRAS.467...50N} that the sizes of the atomic regions (emitting {[\ion{C}{II}]}) shrink with increasing SFR, which can explain the {[\ion{C}{II}]} deficit. At a given SFR, the radius decreases with increasing $G_{0}^{\prime}$.}
 \label{fig:GMCShrink}
\end{figure}

A natural mechanism for the {[\ion{C}{II}]} deficit was presented by \citet{2017MNRAS.467...50N}, where the increase in the molecular fraction of the gas reduces the efficiency of {[\ion{C}{II}]} emission due to the shrinking size of atomic gas in galaxies with high SFR. In Fig.~\ref{fig:GMCShrink}, we present the average radius of the atomic region ($R_{\mathrm{atomic}}$) in the simulated galaxies with respect to the total SFR, colour-coded by the average strength of the radiation field incident on the neutral cloud ($G_{0,\mathrm{cloud}}^{\prime}$). The plot shows a gradual decrease in the effective atomic radius with SFR, which can be related to $G_{0,\mathrm{cloud}}^{\prime}$. $R_{\mathrm{atomic}}$ shrinks because the sizes of the neutral clouds are reduced with increasing SFR, while $R_{\mathrm{H}_2}$ increases. This explains the trends observed in Fig.~\ref{fig:ISMPha}, where the molecular fractional contribution rises after the peak contribution of the atomic gas. 

This result is not surprising as the SFR in the \textsc{EAGLE} simulations is determined by the pressure in the galaxy, and at the same time, pressure constrains the size of the neutral cloud, so these results reflect those of \citet{2017MNRAS.467...50N}. Unfortunately, the evolution of $R_{\mathrm{atomic}}$ cannot be tested at higher SFR for the local Universe in \textsc{EAGLE} \citep{2017MNRAS.472..919K}, but  $R_{\mathrm{atomic}}$ is expected to decrease for systems with higher SFR. 

\begin{figure}
 \includegraphics[width=\columnwidth]{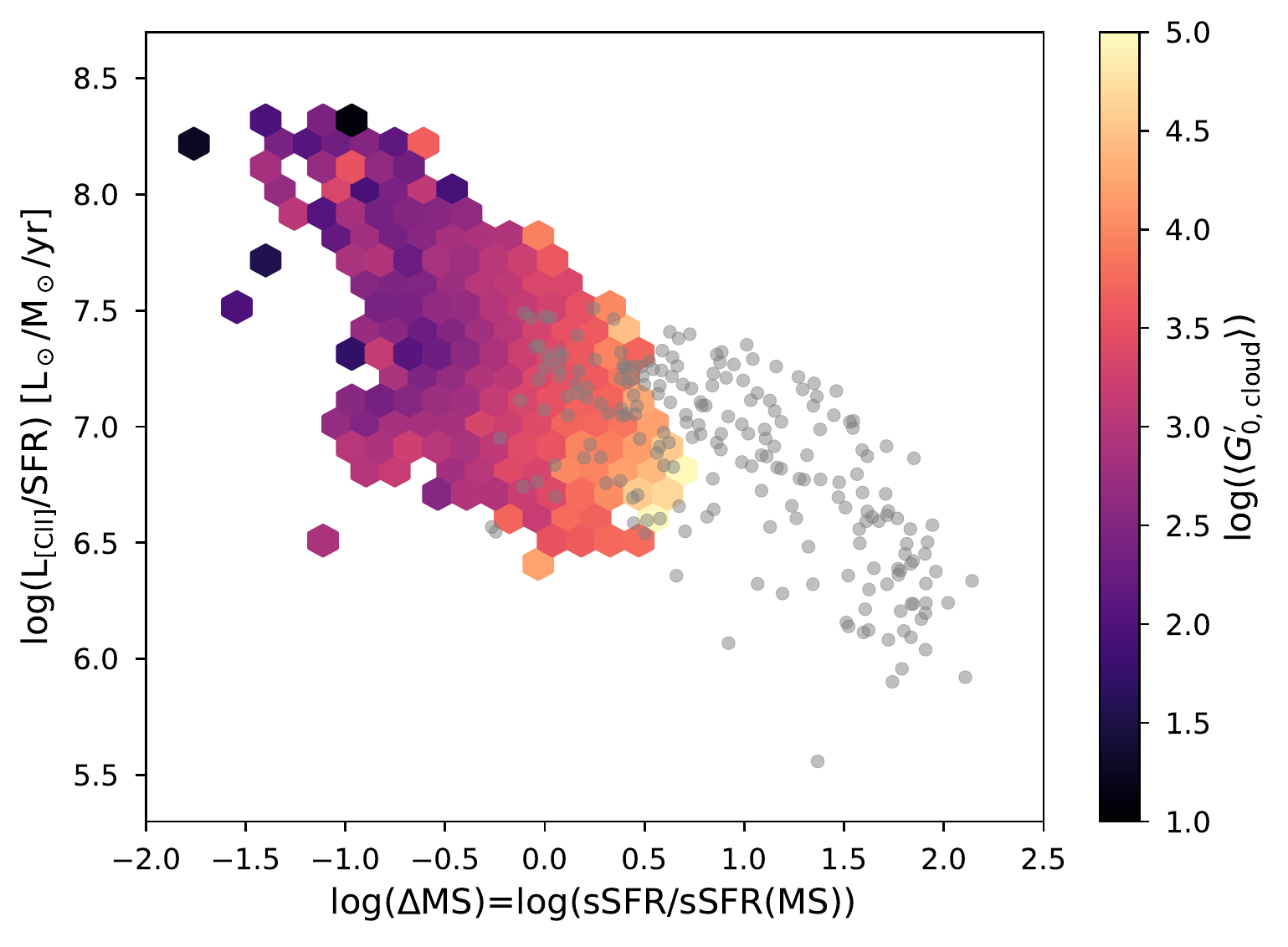}
 \caption{$L_{\mathrm[\ion{C}{II}]}$/SFR ratio as a function of $\Delta$MS for the three simulations (assuming they represent the same population), colour-coded by $G_{0}^{\prime}$ (Habing units). We binned the $L_{\mathrm[\ion{C}{II}]}$/SFR $\Delta$MS values to sizes of $\sim0.1$ dex. When we compare our results with \citet[their Fig.~6]{2018ApJ...861...95H} (grey dots), it is clear that \textsc{EAGLE} simulations do not recover starburst galaxies with $\log{\Delta}$MS $>0.5$, but there is a clear indication that deviations from the MS affects $L_{\mathrm[\ion{C}{II}]}$/SFR.}
 \label{fig:MSDeficit}
\end{figure}

Another claim related to the variations in the $L_{\mathrm[\ion{C}{II}]}$--SFR relation is that $L_{\mathrm[\ion{C}{II}]}$ may not be a robust SFR indicator when intense radiation fields are present, such as in starburst galaxies \citep{2018ApJ...861...95H,2019MNRAS.489....1F}. We test this hypothesis, following \citet{2018ApJ...861...95H}, by calculating the specific star formation rate ($\mathrm{sSFR}=\mathrm{SFR}/\mathrm{M}_{\star}$) for the galaxies in the simulations and normalising with the value derived for the main-sequence (MS) sSFR from \citet{2014ApJS..214...15S}. Figure~\ref{fig:MSDeficit} shows the $L_{\mathrm[\ion{C}{II}]}$/SFR ratio with respect to $\Delta$MS. This result is similar to that presented by \citet[their Fig.~6]{2018ApJ...861...95H} but lacking the starburst outliers ($\Delta\mathrm{MS}\gtrsim20$), which are not reproduced by \textsc{EAGLE}. The $R_{\mathrm{atomic}}$ reduction and the decrease in $L_{\mathrm[\ion{C}{II}]}$/SFR ratio at higher $\Delta$MS show that the strength of the radiation field can be a major factor in the observed variations in the $L_{\mathrm[\ion{C}{II}]}$--SFR relation.

\citet{2017ApJ...846...32D} show that less prominent {[\ion{C}{II}]} deficits are related to higher {[\ion{C}{II}]} fractional contributions coming from the neutral atomic phase. In \citet{2019ApJ...886...60S} the main contributor to $L_{\mathrm[\ion{C}{II}]}$ is found to be the neutral phase, which shows a less prominent {[\ion{C}{II}]} deficit compared to the ionised phase. These results agree with those presented in this work. When the fractional contribution to $L_{\mathrm[\ion{C}{II}]}$ from the atomic phase is small, the deficit is more prominent. The {[\ion{C}{II}]} atomic phase contribution decreases at the same time that the total {[\ion{C}{II}]} luminosity decreases. Thus the neutral phase is responsible for the deficit, as we have shown in this work.

However, when \citet{2019ApJ...886...60S} consider only the ionised component, the {[\ion{C}{II}]} deficit is more prominent compared to the neutral phases. This result is different from what we found in this work, where the {[\ion{C}{II}]} luminosity from the DIG component does not decrease at the higher luminosities ($\log \rm{SFR} \approx 0.8 \sim \log L_{\rm{TIR}} \approx 10.75$ ) tested in this work. The methods used to calculate the ionised component may explain the differences between this work and \citet{2019ApJ...886...60S}. The \citet{2019ApJ...886...60S} calculations come from directly scaling the {[\ion{N}{II}]} $205\,\mu\mathrm{m}$ line deficit, while we calibrate the ionised component taking the two nitrogen emission lines and the SFR into account. The role that metallicity plays in the variation of the $L_{\mathrm[\ion{C}{II}]}$--SFR relation may also explain the differences. The sample of galaxies used by \citet{2019ApJ...886...60S} has, in general, metallicities below solar. Our results (Fig.~\ref{fig:ZEvol}) show that in this range the atomic neutral phase dominates and the DIG contributes less at lower metallicities. In other words, this difference could be due to selection bias. 

\citet{2017ApJ...834....5S} found a correlation between increased metallicity and deeper $L_{\mathrm[\ion{C}{II}]}$ deficits in high-luminosity infrared galaxies. In our case, at higher metallicities, the atomic and molecular gas contributions to the $L_{\mathrm[\ion{C}{II}]}$ seem to be more affected than the DIG contribution (Fig.~\ref{fig:ZEvol}), showing more prominent $L_{\mathrm[\ion{C}{II}]}$ deficits. At values below solar metallicity, the structure of the ISM changes, especially the transition from WNM-to-CNM \citep{2019ApJ...881..160B}. In this work, at $Z_\sun <$ 1 the DIG seems to be deficient in producing {[\ion{C}{II}]} compared to the atomic gas, but the DIG total contribution is always stable. As discussed in the previous paragraph, galaxies in the work by \citet{2019ApJ...886...60S} and \citet{2017ApJ...846...32D} are biased towards metallicities below solar, where atomic phases dominate $L_{\mathrm[\ion{C}{II}]}$. 

The contribution of a given phase to the variations in $L_{\mathrm[\ion{C}{II}]}$--SFR relation depends on the star-formation regulation, due to AGN or star-formation feedback, and metallicity, and these are the keys to understanding the variations observed in line observations. However, we keep in mind that the observed variations in the $L_{\mathrm[\ion{C}{II}]}$--SFR relation can come from selection effects and other biases, such as galaxy brightness, where starburst systems are more easily observed due to their luminosity \citep{2019MNRAS.487.5902K}. Nonetheless, trends in theoretical models are also dependent on the assumed parameters to estimate line emission \citep[e.g.][]{2015ApJ...814...76O,2017ApJ...846..105O,2013MNRAS.433.1567V,2015ApJ...813...36V,2017MNRAS.467.1300V,2018A&A...609A.130L,2019A&A...626A..23C,2019MNRAS.487.5902K,2019MNRAS.482.4906P}. We discuss the caveats of our choices in the following section.  

\subsection{Caveats on our predictions} \label{subsec:DisCaveats}

\subsubsection{Model assumptions}

Our findings may be limited by the assumptions we made in Sect.~\ref{subsec:ISMModel}. We use the model presented by \citet{2015ApJ...814...76O,2017ApJ...846..105O} as a basis. However, we have important differences: We calculate the fraction of neutral hydrogen following \citet{2013MNRAS.430.2427R}, we change the density distribution in the neutral clouds to a Plummer profile following \citet{2019MNRAS.482.4906P} and we calibrate the DIG using {[\ion{N}{II}]} emission lines. The main weakness of our model is its geometric simplicity. The model is based on patches of gas (SPH particles) with assumed constant physical parameters, such as the metallicity, which limits comparisons with real galaxies, where the different ISM phases are entangled with each other \citep{2018Galax...6..100O}. Recent observational results \citep{2019NatAs...3.1114D} have shown that the neutral cloud mass function (Eq.~\ref{eq:GMClaw}) is only valid in the local Universe but not at different redshifts, where neutral clouds could be more massive (higher surface densities). Fortunately, even though this simple approach does not take all galactic physics into account, the models reproduce observed relations in galaxies in the local Universe. 

\citet{2018ApJ...857..148O} apply a geometrical correction to their predicted intensities because of the assumed spherical clouds in the model compared with the computed plane-parallel geometry coming from \textsc{Cloudy}. We do not implement this correction because the effect of this correction is much smaller than the general scatter presented in Fig.~\ref{fig:z0}; therefore, the computational effort of applying the correction is not justified. 

\subsubsection{DIG calibration}

We calibrate the DIG emission (Sect.~\ref{subsec:DIGModel}) to account for the luminosity coming from this ISM phase in the {[\ion{C}{II}]} line. This approach is biased towards the physical properties of the observations used in the calibration, for example luminous galaxies. At the same time, the assumed distribution function for $R_{\mathrm{DIG}}$ (Eq.~\ref{eq:RDIG}) is not entirely physically motivated, as we do not know the actual distribution of this ISM phase in different types of galaxies. As we show in Fig.~\ref{fig:ISMPha}, this calibration can affect the DIG contribution of the {[\ion{C}{II}]} line. The spatial distribution  presented in Fig.~\ref{fig:Radial} could also be affected by the uncertainty of $R_{\mathrm{DIG}}$. With smaller (bigger) $R_{\mathrm{DIG}}$ the density of the ionised clouds will increase (decrease), the gas becomes less (more) diffuse, thus the total luminosity coming from the DIG will increase (decrease) as well. However, the assumptions used in this work can give us insights into the general behaviour of the DIG (see the discussion in Sect.~\ref{subsec:DisRightSim}). 
Different physical assumptions for the DIG could be applied to the simulations, but most of them only assume the distribution of the ionised gas in disky galaxies \citep[e.g.][and references therein]{2009RvMP...81..969H,2019MNRAS.488.1977V}. Models with reliably calibrated DIG properties would be very important to understand diagnostic line emissions in galaxy evolution \citep{2019ARA&A..57..511K}, and more observations focusing on this ISM phase are required, such as the GBT Diffuse Ionised Gas Survey\footnote{\url{https://greenbankobservatory.org/science/gbt-surveys/gdigs/}}.

In this work, we present luminosity predictions coming from {[\ion{C}{II}]} emission line with calibrated emission from {[\ion{N}{II}]} emission lines, which is ideal in terms of modelling \citep{2018Galax...6..100O}, as other lines can help to constrain the physical properties in the models to improve emission line predictions \citep{2019MNRAS.482.4906P}. We plan to study the local Universe and at high redshift with more FIR emission lines in future works. 

\subsection{Choice of simulation}\label{subsec:DisRightSim}

The findings of this work may be somewhat limited by the selection of a given \textsc{EAGLE} simulation to estimate the line emission. For example, large cosmological volumes ($L>50\,$cMpc) provide a more representative range of environments in galaxies compared to small boxes \citep{2015MNRAS.446..521S,2015MNRAS.450.4486F}. In addition, the efficiency of feedback must be calibrated (e.g. with the GSMF) in large hydrodynamical simulations where the ISM scales are unresolved, and therefore calibration is required to make valid predictions for observables \citep{2015MNRAS.446..521S,2017ARA&A..55...59N}.

We have presented results from three simulations (\textsc{Ref-L0025N0376}, \textsc{Recal-L0025N0752} and \textsc{Ref-L0100N1504}) to test the effects of the box size and resolution on the predictions. We presented some of these comparisons in Figs.~\ref{fig:ISMPha}--\ref{fig:ZEvol}. In terms of the simulated box size (\textsc{Ref-L0025N0376} vs \textsc{Ref-L0100N1504}), the predictions are very similar, and only the total number, typical masses and SFR of galaxies change. As remarked above, \textsc{EAGLE} does not contain many galaxies with SFR $>1 \,\mathrm{M_{\sun}}\,\mathrm{yr^{-1}}$ at $z=0$ due to the specific implementations \citep{2017MNRAS.472..919K,2019A&A...624A..98W}; therefore, with \textsc{EAGLE} simulations we can predict line emissions for normal star-forming galaxies at $z=0$, but not starburst-like systems.  

The trends in Figs.~\ref{fig:z0} and \ref{fig:ZEvol} show that $L_{\mathrm[\ion{C}{II}]}$ is always lower for \textsc{Recal-L0025N0752} galaxies. These lower values are due to the GSMF re-calibration. \citet{2017MNRAS.472.3354D} noted that the increase in the resolution for these simulations (\textsc{Ref-L0100N1504} to \textsc{Recal-L0025N0752}) can affect some features but the fundamental metallicity scaling relations are not altered. Therefore, boxes re-calibrated to the GSMF are a good starting point to predicting emission lines as metallicity scaling relations holds. From our comparison with observations (Sect.~\ref{subsec:ISMPhase}), we expect that the observed galaxies in the local Universe behave as in \textsc{Recal-L0025N0752}, where the atomic gas dominates at SFR values higher than $0.01\,\mathrm{M_{\sun}}\,\mathrm{yr^{-1}}$. 

\textsc{Ref-L0100N1504} provides (statistical) clues on how the fractional contribution of the ISM phases changes. Predicted luminosities from \textsc{Ref-L0100N1504} and \textsc{Ref-L0025N0376} are overestimated compared with observed galaxies, which are well reproduced by \textsc{Recal-L0025N0752}. This conclusion can also be seen from the right panel of Fig.~\ref{fig:ZEvol}, where a reduction of the contribution from atomic and molecular phases in $L_{\mathrm[\ion{C}{II}]}$ increases the dominance of the DIG phase. To summarise, \textsc{Recal-L0025N0752} gives us the best predictions in terms of reproducing observed local galaxies and \textsc{Ref-L0100N1504} gives us larger sample sizes for statistical studies.

As \textsc{Recal-L0025N0752} does not reproduce the large number of galaxies needed for some statistical studies, we suggest that the best way to compare our simulations to observations is a mix between \textsc{Ref-L0100N1504} and \textsc{Recal-L0025N0752} to observe the global behaviour of the emission lines, as we do in Figs.~\ref{fig:GMCShrink} and \ref{fig:MSDeficit}. This gives us a better idea of the contribution of the different ISM phases to the line emission.

\section{Summary and conclusions} \label{sec:conclu}

In this work, we have post-processed SPH \textsc{EAGLE} simulations using a multi-phase ISM model and then predicted the luminosity of {[\ion{C}{II}]} emission at 158 $\mu$m with \textsc{Cloudy}. We set out to determine the fractional contributions of the different ISM phases to the {[\ion{C}{II}]} line and the effect of these phases on the $L_{\mathrm[\ion{C}{II}]}$--SFR relation. We use three sets of simulations (\textsc{Ref-L0025N0376}, \textsc{Recal-L0025N0752} and \textsc{Ref-L0100N1504}) with two \textsc{Cloudy} cooling tables for shielded and optically thin regimes \citep{2020MNRAS.497.4857P} to characterise ISM composition in terms of dense molecular gas, neutral atomic gas, and diffuse ionised gas (DIG). We validate our model by comparing with observations of local galaxies. Our main conclusions are the following: 

   \begin{enumerate}
    \item We find a dependence of the fractional contribution of the different ISM phases to $L_{\mathrm[\ion{C}{II}]}$ on the total SFR and metallicity. Our model agrees with previous works where the {[\ion{C}{II}]} emission comes mainly from neutral ISM regions in observed galaxies. However, this could be a selection bias in the observations towards galaxies with metallicities below solar. 
      \item In systems where the SFR is low, the DIG plays a dominant role in producing $L_{\mathrm[\ion{C}{II}]}$. Additional resolved observations of local galaxies with low SFRs and their outskirts will be essential to improve the calibration of the DIG fraction in other emission lines \citep{2018PASA...35....2V}.
      \item The model for the ISM we have implemented in this work reproduces the observed $L_{\mathrm[\ion{C}{II}]}$--SFR relation in local galaxies, although this result depends on calibrating the DIG with {[\ion{N}{II}]} emission lines. 
      \item Star-formation regulation and metallicity dependence of the different ISM phases could be responsible for the variations observed in $L_{\mathrm[\ion{C}{II}]}$ at high infrared luminosities, based on the dependence on $\Delta$MS and average $R_{\mathrm{cloud}}$. Further investigations are needed to verify if this result holds at higher redshifts.
      \item The use of large boxes (box-size $\gtrsim 100$ cMpc) and high-resolution simulations (mass resolution $\lesssim10^{5}\,\rm{M}_{\sun}$) is key to correctly predict emission lines in different types of galaxies.
   \end{enumerate}
   
In the future, we will take black hole particles (especially the SMBH) into account, which will alter the radiation field in the centres of the galaxies. We will use a random sample of galaxies from the simulations to estimate other fine structure lines. We plan to compare predictions in high-redshift systems with observational results, such as those presented by \citet{2019arXiv191009517L} and \citet{2019ApJ...882...10N} with a large number of observed galaxies, using for example $\Delta$MS.

\begin{acknowledgements}
The authors thank Rodrigo Herrera-Camus for providing the data from the SHINING survey. We acknowledge the anonymous referee for a careful reading of the manuscript and very helpful questions and comments. We acknowledge the Virgo Consortium for making their simulation data available. The EAGLE simulations were performed using the DiRAC-2 facility at Durham, managed by the ICC, and the PRACE facility Curie based in France at TGCC, CEA, Bruy\`{e}res-le-Ch\^{a}tel. This research made use of Astropy,\footnote{http://www.astropy.org} a community-developed core Python package for Astronomy \citep{2013A&A...558A..33A,2018AJ....156..123A}. We would like to thank the Center for Information Technology of the University of Groningen for their support 
and for providing access to the Peregrine high performance computing cluster.
\end{acknowledgements}

\bibliographystyle{aa}
\bibliography{main}

\begin{appendix}

\section{Comments on archival Data}\label{App:Observations}

To complement and verify our model in the local Universe, we use archival data from different samples of observed galaxies where the {[\ion{C}{II}]} luminosity is available. In cases where it is possible, we recalculate the luminosities and SFRs with the same cosmology used in this work \citep{2014A&A...571A...1P}. We use the samples from (a) the ISO compendium \citep{2008ApJS..178..280B} of main-sequence galaxies; (b) xCOLD GASS \citep{2017MNRAS.470.4750A}, composed of intermediate-stellar-mass galaxies; (c) GOALS \citep{2013ApJ...774...68D,2017ApJ...846...32D}, composed of LIRGS observed with \textit{Spitzer} and \textit{Herschel}; (d) SHINING \citep{2018ApJ...861...95H}, composed of nearby star-forming galaxies, AGNs, and LIRGs observed with \textit{Herschel}; (e) VALES \citep{2017A&A...602A..49H}, composed of dusty main-sequence star-forming galaxies observed with \textit{ALMA}; and (f) the Herschel Dwarf Galaxy Survey \citep[HDGS,][]{2015A&A...578A..53C,2019A&A...626A..23C}. For the ISO compendium: We ignore close galaxies (below 1 Mpc), we use the median for the line flux in galaxies with more than one measure in the {[\ion{C}{II}]} line, and the infrared luminosity was calculated with the 60 and 100 $\mu$m from the IRAS flux as described by \citet{2008ApJS..178..280B}. In most of the literature samples we have calculated the SFR from the IR luminosity as described by \citet{2012ARA&A..50..531K}. We use the SFRs from \citet{2017MNRAS.470.3775V} and \citet{2016MNRAS.462.1749S} for 
VALES and xCOLD GASS samples, respectively.

\begin{figure}
 \includegraphics[width=\columnwidth]{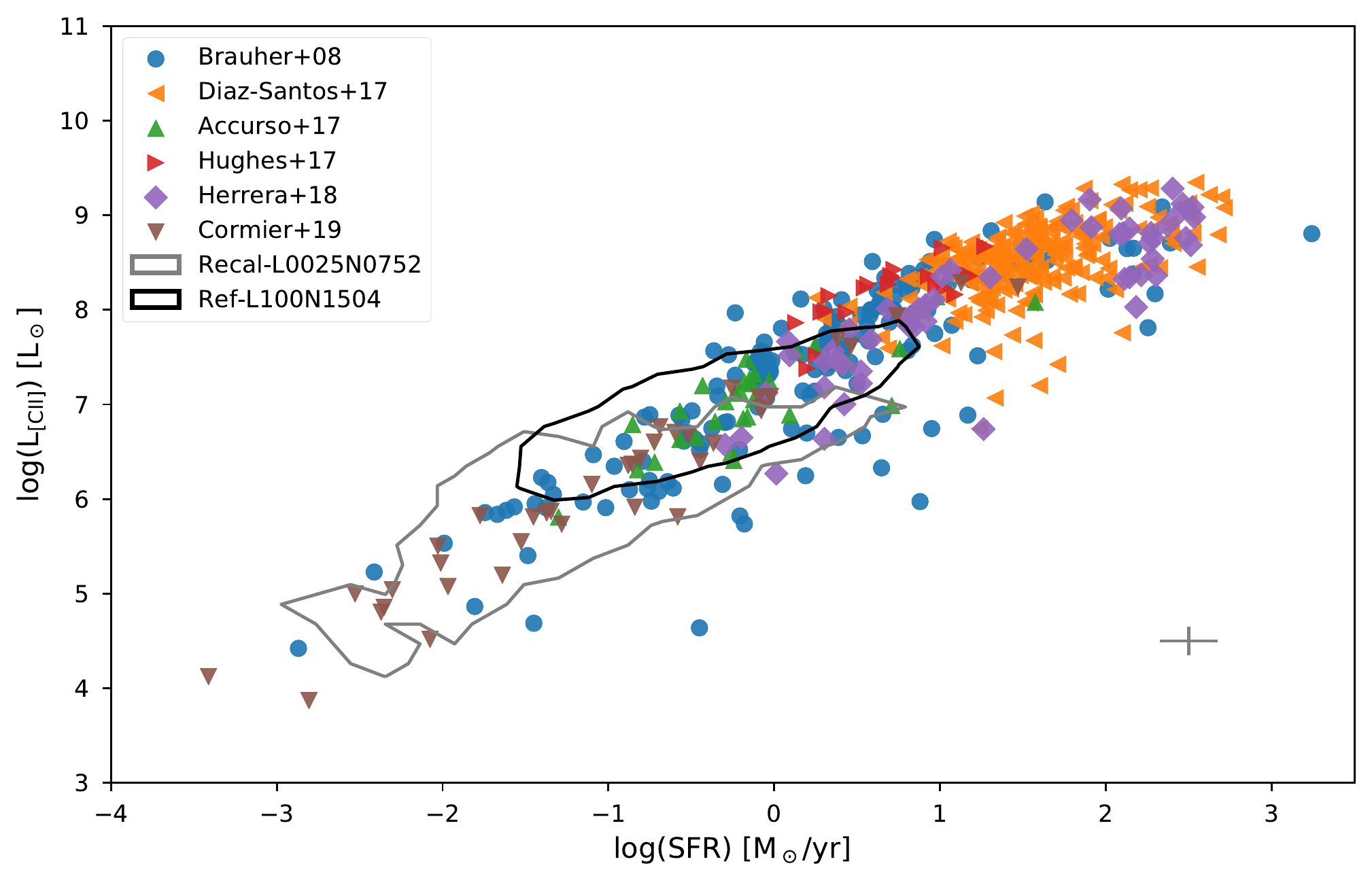}
 \caption{$L_{\mathrm[\ion{C}{II}]}$--SFR relation for the observational sample of local galaxies and two simulations used in this work (\textsc{Recal-L0025N0752}, \textsc{Ref-L0100N1504}) presented as contour maps (grey and black). The contours shows where 95\% of the galaxies of the respective simulations fall in the relation. We present the mean error from the observational samples in the bottom-right corner of the plot.}
 \label{fig:z0Appendix}
\end{figure}

We compare these literature samples with our SFR and $L_{\mathrm[\ion{C}{II}]}$ estimates in Fig.~\ref{fig:z0Appendix}. We notice that galaxies in \textsc{Recal-L0025N0752} follow similar trend as the dwarf galaxies from \citet{2019A&A...626A..23C} while \textsc{Ref-L0100N1504} is similar to most of the intermediate stellar mass galaxies from \citet{2017MNRAS.470.4750A}. Both simulations reproduce main sequence galaxies \citep{2008ApJS..178..280B} as well as some galaxies from other samples \citep[e.g.][]{2017A&A...602A..49H,2018ApJ...861...95H}. LIRGS galaxies \citep{2017ApJ...846...32D,2018ApJ...861...95H} are representative of the {[\ion{C}{II}]} deficit, but unfortunately,  are not recovered by EAGLE (see Sect.~\ref{subsec:CII-SFR} and Fig.~\ref{fig:z0} to compare with the deficit). 

Additionally, we use {[\ion{N}{II}]} observational data from \citet{2016ApJS..226...19F} to calibrate the DIG. We describe the procedure in Sect.~\ref{subsec:DIGModel} and in Fig.~\ref{fig:z0NII} we show the simulated and observed $L_{\mathrm[\ion{N}{II}]}$--SFR relation for the lines at 122 and 205 $\mu$m. 

\end{appendix}

\end{document}